\documentclass[prc,amsmath,amssymb,twocolumn,floatfix, nofootinbib]{revtex4-1}

\usepackage{graphicx,bm,subfigure}
\usepackage{color}

\graphicspath{{figures/}}

\newcommand{\beq}{\begin{equation}}
\newcommand{\eeq}{\end{equation}}
\newcommand{\bea}{\begin{eqnarray}}
\newcommand{\eea}{\end{eqnarray}}
\newcommand{\bi}{\begin{itemize}}
\newcommand{\ei}{\end{itemize}}
\newcommand{\ben}{\begin{enumerate}}
\newcommand{\een}{\end{enumerate}}

\newcommand{\comm}[2]{\ensuremath{[#1,#2]}}

\newcommand{\bra}[1]{\ensuremath{\langle#1|}}
\newcommand{\ket}[1]{\ensuremath{|#1\rangle}}

\newcommand{\Hdiag}{\ensuremath{H_{d}}}  
\newcommand{\Trel}{\ensuremath{T_{\rm rel}}}

\newcommand{\fmi}{\, \text{fm}^{-1}}
\newcommand{\mev}{\, \text{MeV}}

\newcommand{\km}{\ensuremath{{k^{\protect\phantom{\prime}}}}}
\newcommand{\kpm}{\ensuremath{{k^{\prime}}}}

\newcommand{\LambdaEFT}{\Lambda_{\rm EFT}}

\newlength{\wideplotwidth}
\newlength{\movieplotwidth}
\newlength{\phaseplotwidth}
\newlength{\universalityplotwidth}
\begin{document}

\title{Decoupling of Spurious Deep Bound States \\ 
       with the Similarity Renormalization Group}

\author{K.A.\ Wendt}
\email{wendt.31@osu.edu}
\affiliation{Department of Physics, The Ohio State University, Columbus, OH 43210}
\author{R.J.\ Furnstahl}
\email{furnstahl.1@osu.edu}
\affiliation{Department of Physics, The Ohio State University, Columbus, OH 43210}
\author{R.J.\ Perry}
\email{perry@mps.ohio-state.edu}
\affiliation{Department of Physics, The Ohio State University, Columbus, OH 43210}

\date{\today}

\begin{abstract}
The Similarity Renormalization Group (SRG) is a continuous series
of unitary transformations that can be implemented as a flow
equation.  When the relative kinetic energy ($\Trel$) is
used in the SRG generator, nuclear structure calculations have shown
greatly improved convergence with basis size because of the decoupling
of high-energy and low-energy physics. However this generator can 
sometimes be problematic.
A test case is provided by a study of initial interactions from 
chiral effective field theories with large cutoffs, which can lead to
spurious deep bound states.  We would like the SRG to decouple these
from the physical shallow bound states. However, with $\Trel$ 
the high- and low-energy bound states are not decoupled in the
usual sense.  Replacing $\Trel$ by the momentum-space diagonal of the
Hamiltonian ($H_d$) in the SRG generator does produce
decoupling, such that the shallow states are in the low-momentum
region and the deep bound states are at higher momentum. 
The flow toward universal low-momentum interactions
is also restored. 

\end{abstract}

\smallskip
\newpage
\maketitle


\section{Introduction}
\label{sec:intro}
    
Similarity Renormalization Group (SRG) flow equations with a generator
based on the relative kinetic energy have been shown to soften
inter-nucleon interactions by decoupling energy 
scales~\cite{Jurgenson:2007td,Bogner:2009bt},  leading to
significantly enhanced convergence in calculations of nuclear
structure and 
reactions~\cite{Bogner:2007rx,Jurgenson:2009qs,Jurgenson:2010wy,Hebeler:2010xb,Bogner:2009bt,
Navratil:2010jn,Navratil:2010ey}.
To date, these SRG applications have all started with phenomenological
potentials such as Argonne $v_{18}$~\cite{Wiringa:1994wb} or the
N$^3$LO  chiral potentials from Refs.~\cite{Entem:2003ft} and
\cite{Epelbaum:2004fk}.
When evolved to softened form, matrix elements of these interactions
are found to collapse toward a common NN potential for momenta below
the decoupling scale~\cite{Bogner:2009bt}.    
These desirable SRG features are not guaranteed, however, when applied
to chiral effective field theory (EFT) interactions with large
cutoffs.   Such interactions have been explored at leading
order~\cite{Nogga:2005hy} to examine renormalization  issues (see also
Refs.~\cite{PavonValderrama:2005wv,PavonValderrama:2005gu,
PavonValderrama:2005uj,
Valderrama:2009ei,Yang:2009fm,Machleidt:2009bh, Phillips:2010rv}).
The SRG approach as applied so far to nuclear interactions fails for
these theories in channels where the tensor force from pion exchange
introduces spurious deep bound states.
The generic cause of this failure was identified by Glazek and
Perry~\cite{Glazek:2008pg}. In this paper we document the specific
problems with large-cutoff chiral EFT ($\chi_{\rm EFT}$) and test whether they are fixed
as described in Ref.~\cite{Glazek:2008pg}.    
    
The SRG as used here is implemented as a series of infinitesimal
unitary transforms of a Hamiltonian $H$, labeled by a flow parameter
$s$~\cite{Wegner:1994,Kehrein:2006},
\beq
 H(s) =  U(s) H(s=0) U^\dagger(s)
 \;.
\eeq
Differentiating with respect to the flow parameter 
shows how any evolution is constrained by unitarity,
\beq
  \frac{d}{ds}H(s) = \comm{\eta(s)}{H(s)} \;,  
\eeq
where
\beq
  \eta(s) = \frac{d U(s)}{ds}U^\dagger(s)
  \;.
\eeq
The flow is controlled through the choice of the anti-Hermitian
generator $\eta(s)$.  
To generate a renormalization group evolution, we need the flow to decouple high and low energy degrees of freedom (i.e.\ to remove far off diagonal matrix elements).  
For many applications, a useful form for the generator is
\beq
  \eta(s) = \comm{G(s)}{H(s)} \;,
\eeq
where $G(s)$ is a Hermitian operator.  
The original choice for $G(s)$ advocated by Wegner and 
collaborators~\cite{Wegner:1994,Kehrein:2006} and applied extensively
in condensed matter is the diagonal component of the interaction,
$G(s) = \Hdiag(s)$,
\beq
  \bra{i}\Hdiag(s)\ket{j} \equiv
  \begin{cases}
  \bra{i}H(s)\ket{i} & \text{if } i = j \;,\\
    0 & \text{otherwise.}
  \end{cases}
\eeq
If the basis in which $\Hdiag$ is taken to be diagonal is a
discretized, partial-wave momentum basis $\ket{i}\equiv\ket{k_ilm}$,
the Hamiltonian is driven toward diagonal form with $s$ such that the
parameter $\lambda \equiv s^{-{1}/{4}}$ is a measure of the width
of diagonal~\cite{Wegner:1994,Kehrein:2006}.

In most applications to nuclear systems, $G(s)$ has been taken simply
to be the relative kinetic energy $\Trel$, which is by construction
independent of $s$.  This choice also has the effect in practice of
driving the momentum-space potential toward the diagonal, which
directly softens the  repulsive core of nucleon-nucleon interactions
(as well as short-range tensor forces)  by decoupling high- and
low-momentum degrees of freedom~\cite{Jurgenson:2007td}. No adverse
results have been observed for light nuclei; however,  this SRG
evolution has always been terminated such that $\lambda \ge
1\,\fmi$.  

In Ref.~\cite{Glazek:2008pg}, it was observed that when evolving a
simple model Hamiltonian,  the Wegner evolution ($G = \Hdiag$) will
decouple the bound state by leaving it as an isolated eigenvalue on
the diagonal of the Hamiltonian matrix. 
In contrast, they found a very different behavior with the evolution 
generated using $\Trel$.  With $\Trel$, the bound state remains
coupled to low momentum, and is pushed to the lowest momentum part
of the matrix.
This undesirable result has not been seen in the two-nucleon system
for $\lambda \ge 1\,\fmi$ because the only physical bound state, the
deuteron, is very shallow (e.g., the scattering length in this channel
is unnaturally large) and only plays a role for $\lambda \ll 1\,\fmi$.
Indeed, the evolution  for $\Trel$ and $\Hdiag$ in the usual range of
$\lambda$ is practically identical numerically (at least for the
$A=2$ system).

In most cases, we do not want to decrease $\lambda$ to the point where
bound-state eigenvalues appear on the diagonal. 
The effects of discretized momenta are magnified when $H$ is driven
too close to diagonal form, so we typically seek a window in which
high-energy states decouple but the low-energy part of $H$ is
insensitive to discretization.
However, if a deep spurious bound state exists because our effective
Hamiltonian is singular at short distances, we want to make sure that
any associated spurious physics decouples from the low-energy physics
our effective theory is intended to capture. In this way the practical
advantages of the SRG are preserved. This is accomplished if the
spurious bound eigenstate is forced to the diagonal when $\lambda$ is
still sufficiently above the scale of interest. 
According to Glazek and Perry, this happens for the \Hdiag-based generator but $\Trel$ drives       
the bound-state eigenvalue to the low-energy corner of the Hamiltonian
matrix~\cite{Glazek:2008pg}.  As a result, $\Hdiag$ decouples the spurious state from
low-energy physics but $\Trel$ causes the spurious state to corrupt
low-energy physics.                                                  

In this paper, we test whether the analysis of Glazek and Perry
carries over to the nuclear case with spurious bound states
and further investigate the nature of decoupling and
the flow toward universal interactions in the SRG.
In Section~\ref{sec:background}, we give background details on the leading-order
chiral EFT that is the laboratory for our analysis.
In Section~\ref{sec:results}, 
we examine decoupling for the deuteron and phase shifts starting
from the LO EFT.
The implications are discussed in Section~\ref{sec:discussion}
and we summarize in Section~\ref{sec:summary}.


\section{Background}\label{sec:background}

\subsection{Chiral EFT at Leading Order}

An implementation of chiral EFT at leading order (LO) in Weinberg's power counting was studied in
Ref.~\cite{Nogga:2005hy} for a large range of momentum cutoffs
$\LambdaEFT$ to determine what renormalization was needed in each
partial wave to achieve cutoff independence.  The EFT NN potential at
LO consists of one-pion exchange in all channels plus a regulated
contact interaction in certain channels; it is meant to be iterated to
all orders.  
The power counting proposed by Weinberg included contact interactions
in only the S-waves at LO, but this was found by Nogga et al.\ to be
inadequate for cutoff independence in partial waves that have  an
attractive tensor force (for high enough angular momentum there is no
problem).  Adding a single contact interaction in applicable channels
remedied this problem and allowed large cutoffs to be explored.  

A side effect of the attractive tensor force was the appearance with
increasing $\LambdaEFT$ of spurious deep bound states.
These are not a problem in principle for the EFT, because they appear
outside its domain of validity.  That is, the EFT is expected to have
incorrect ultraviolet (UV) behavior, whose impact on low-energy
observables is corrected by the contact
counterterms.  In this case, UV refers to large negative as well as
positive high energies. In practice the problems caused by spurious deep bound states for solving few-body equations led Nogga et al.\
to remove them by hand~\cite{Nogga:2005hy}.
Our question here is: Can these deep bound states be decoupled by
renormalization group flow equations?  This question makes the LO
chiral EFT Hamiltonian a good test case to study the robustness of the
observations in Ref.~\cite{Glazek:2008pg} and to learn about the
decoupling properties of the SRG as a function of $\LambdaEFT$. 
The present study is restricted to two-particle systems.  The  impact
on three- and higher-particle systems may be of even greater interest,
but is deferred to future work.

\begin{figure}[tbh-]
  \includegraphics[width=0.80\columnwidth]{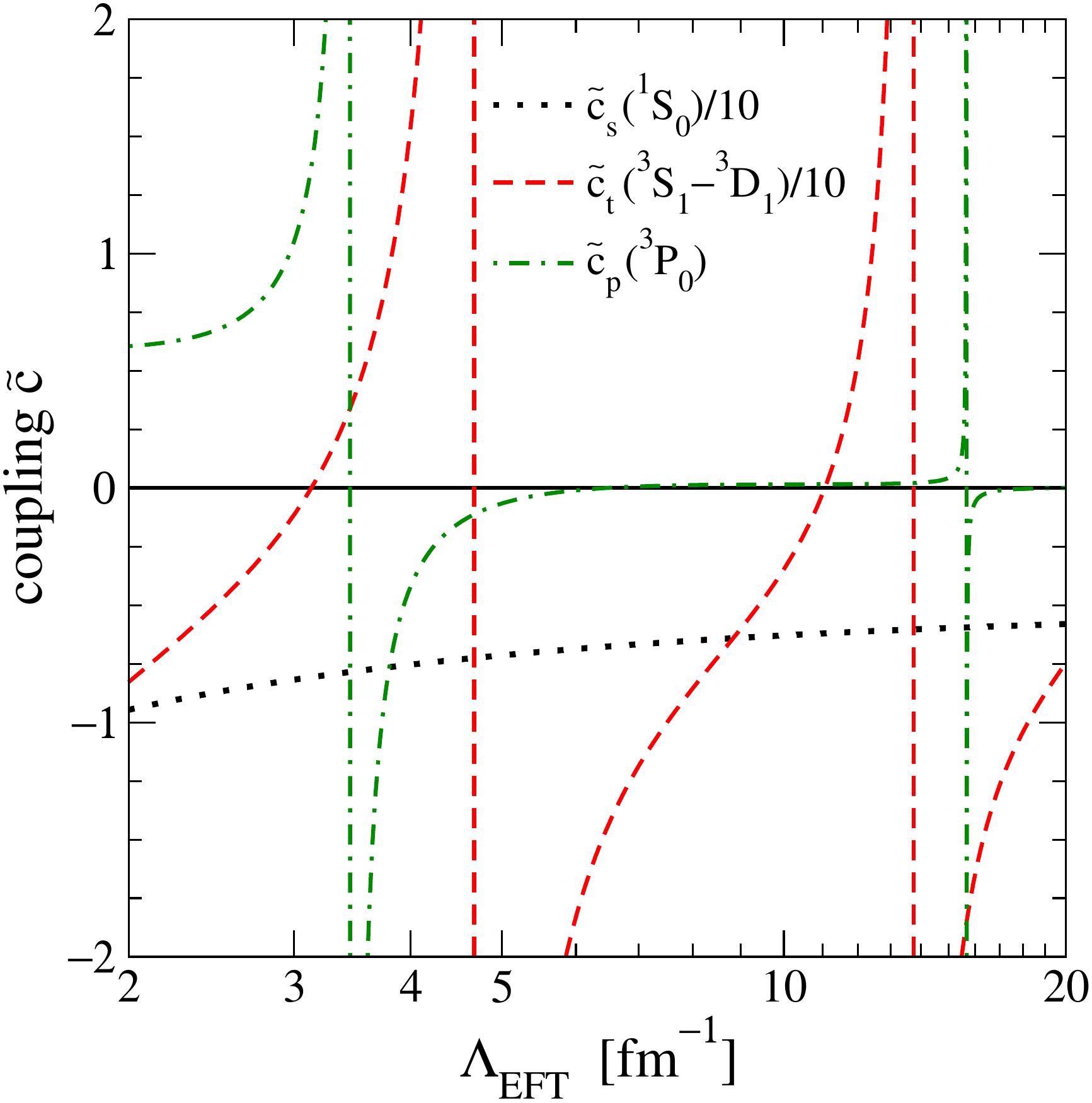}
  \vspace*{-.10in}
  \caption{(color online) Dimensionless coupling constants for the LO
chiral EFT  as a function of the EFT cutoff $\LambdaEFT$. For the
$^1$S$_0$ channel, $\tilde c_s \equiv c_s/f_\pi^2$ is fit to the phase
shift at $\rm E_{\rm lab} = 10\mev$.  For the coupled $^3$S$_1$--$^3$D$_1$ channel,
$\tilde c_t \equiv c_t/f_\pi^2$ is fit to the $^3$S$_1$ phase shift at
$10\mev$. The constant $\tilde c_p \equiv c_p/f_\pi^4$  
is fit to the $^3$P$_0$ phase shift at $50\mev$.
\label{fig:coupling}}
\end{figure}

We follow Ref.~\cite{Nogga:2005hy} in defining the LO chiral
EFT. For all our calculations, we regulated the interaction with
\beq
 f_n(\km,\kpm)=e^{-({\km}/{\LambdaEFT})^{2n}}
  e^{-({\kpm}/{\LambdaEFT})^{2n}} \;,
  \label{eq:regulator}
\eeq
using $n=4$.
Our notation differs from Ref.~\cite{Nogga:2005hy} only in the
definition of dimensionless coupling constants (see Fig.~\ref{fig:coupling} caption). 
We focus on three channels, which are sufficient for the present
discussion.  The $^1$S$_0$ channel is an example of a partial
wave where the tensor force does not contribute and therefore
has no spurious bound states.
The coupled $^3$S$_1$--$^3$D$_1$ channel and the $^3$P$_0$
channel provide representative examples of partial waves with an
attractive tensor force; each develops spurious bound states
with increasing $\LambdaEFT$ in addition to the physical
deuteron in the coupled channel.

\begin{figure}[tbh-]
  \includegraphics[width=0.80\columnwidth]{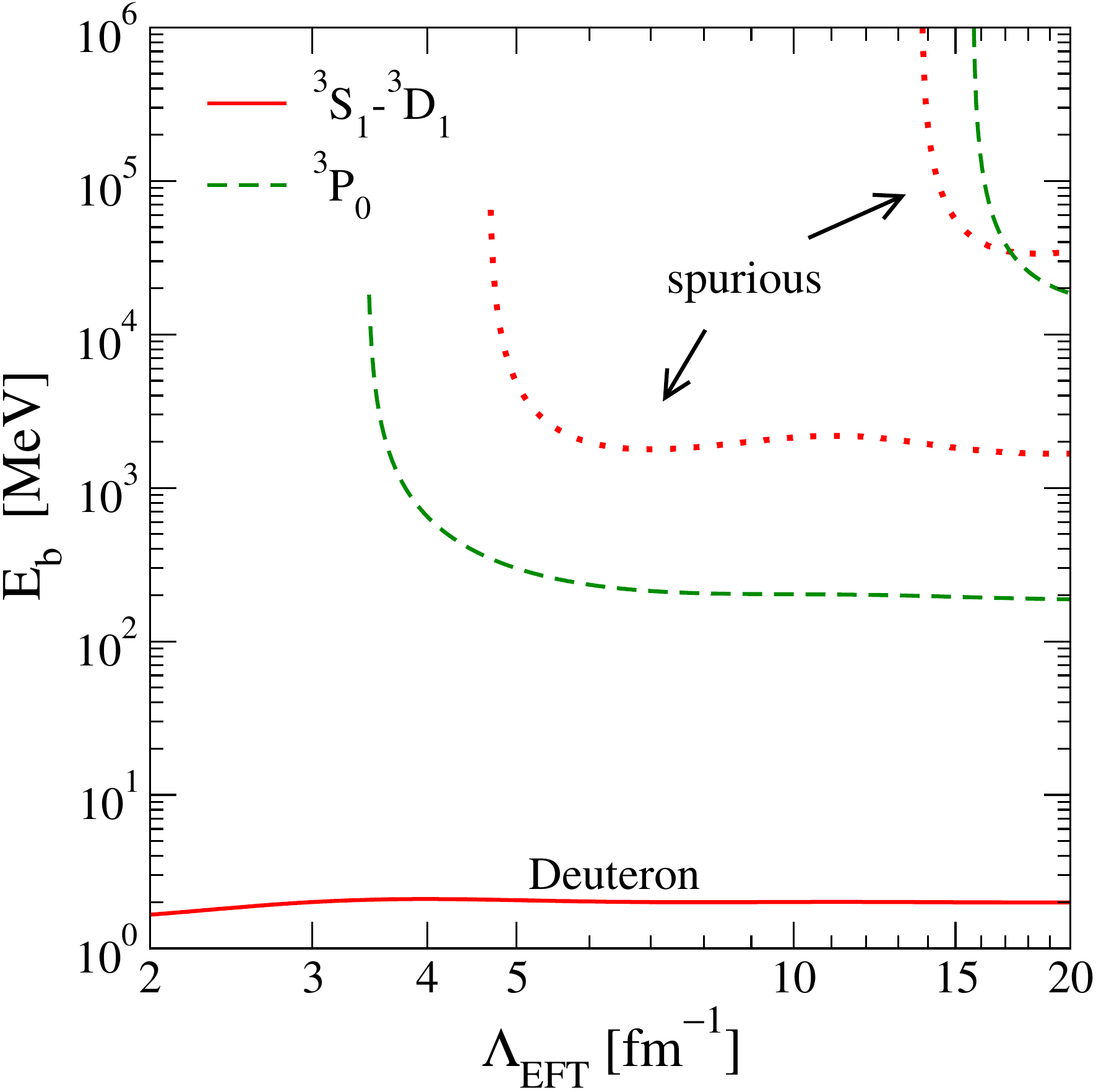}
  \vspace*{-.10in}
  \caption{(color online) Bound-state energies as a function of $\LambdaEFT$
  for the $^3$S$_1$--$^3$D$_1$ coupled channel (deuteron as a solid
  line, spurious states as dotted lines) and the
  $^3$P$_0$ channel (spurious states as dashed lines).}
  \label{fig:BSeft}
\end{figure}

\begin{figure}[tbh-]
  \includegraphics[width=.80\columnwidth]{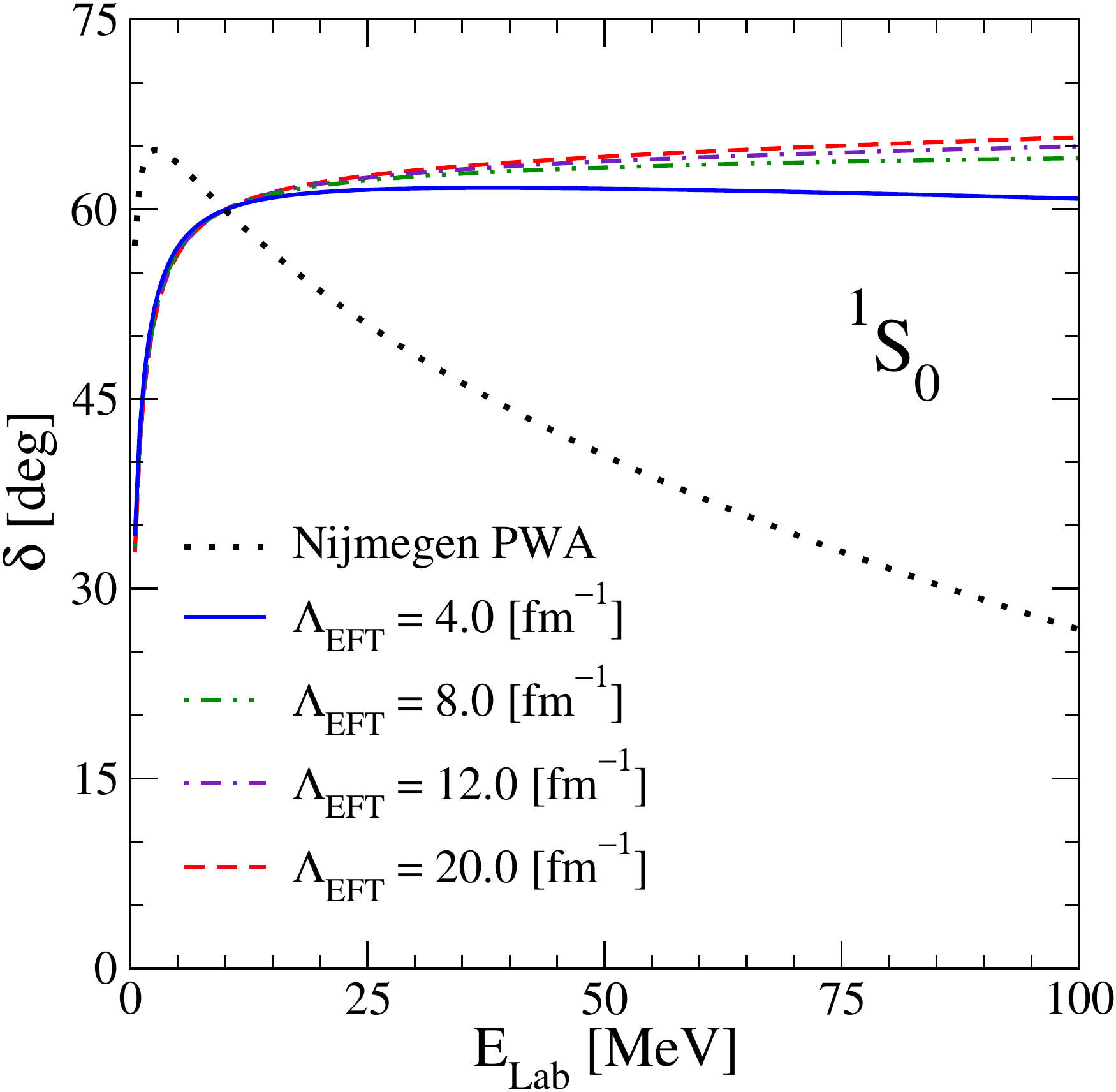}
  \vspace*{-.10in}
  \caption{(color online) $^1$S$_0$ phase-shifts as a function of laboratory energy for several different $\LambdaEFT$.
  \label{fig:1S0EFTPhaseShift}}
\end{figure} 

\begin{figure}[tbh-]
  \includegraphics[width=.80\columnwidth]{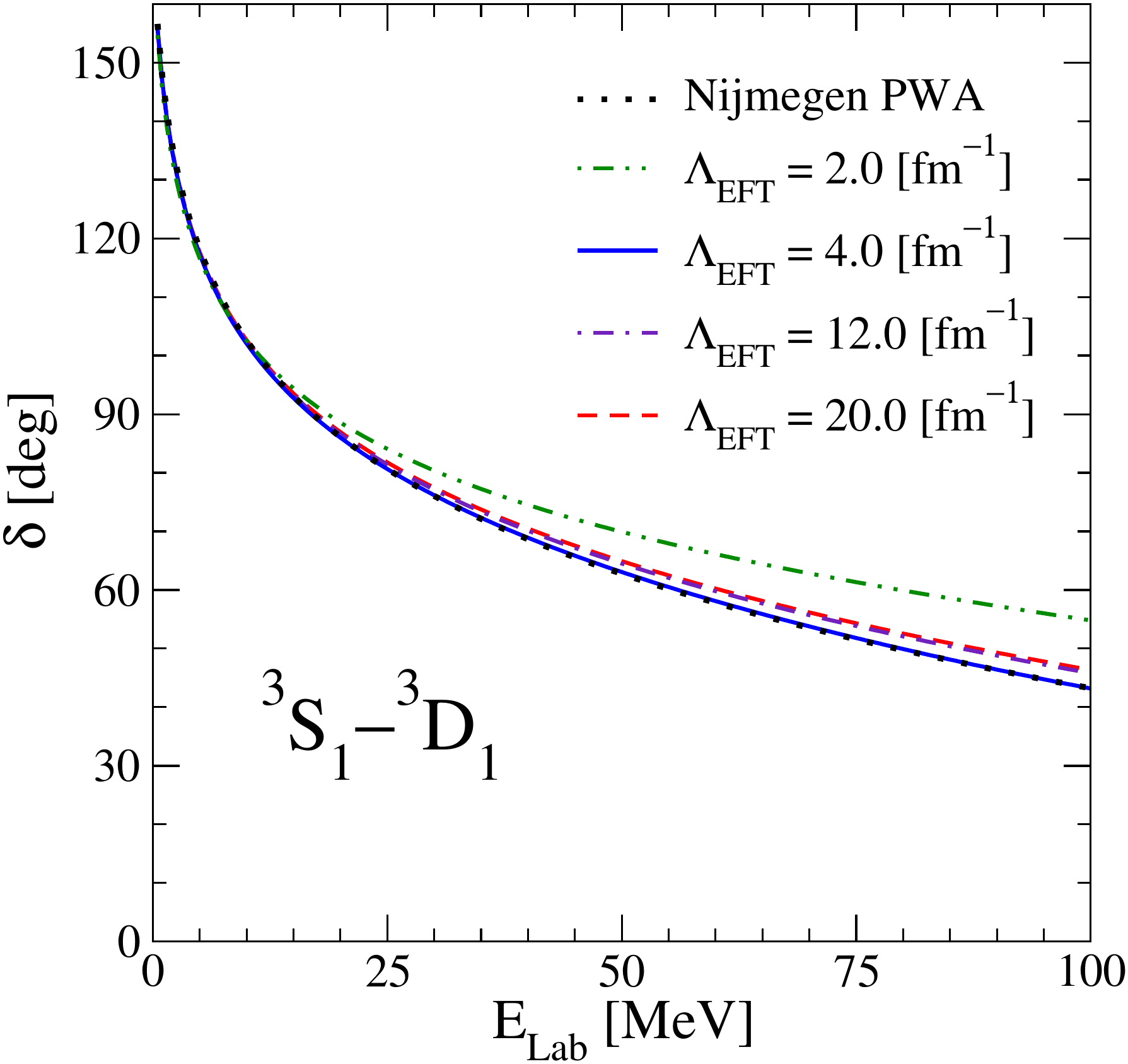}
  \vspace*{-.10in}
  \caption{(color online) $^3$S$_1$ phase-shifts as a function of laboratory energy for several different $\LambdaEFT$.
  \label{fig:3S1EFTPhaseShift}}
\end{figure}  

\begin{figure}[tbh-]
  \includegraphics[width=.80\columnwidth]{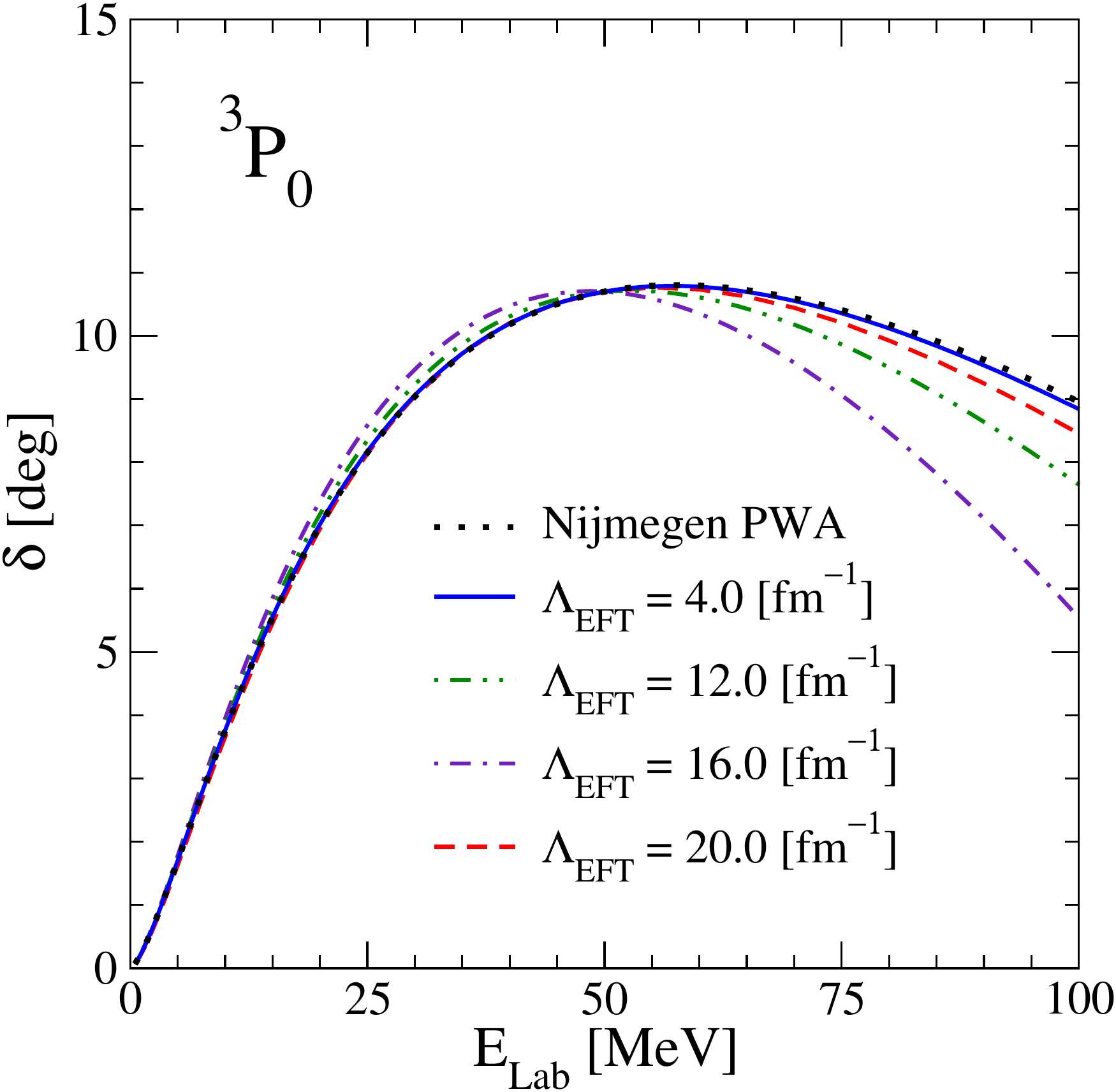}
  \vspace*{-.10in}
  \caption{(color online) $^3$P$_0$ phase-shifts as a function of laboratory energy for several different $\LambdaEFT$.
  \label{fig:3P0EFTPhaseShift}}
\end{figure}

For the
singlet and triplet S-waves we fit the coupling constants
to the neutron-proton phase shifts 
at $E_{\rm lab} = 10\mev$, and for the $^3$P$_0$ channel we fit to the 
$E_{\rm lab} = 50\mev$ phase shift.  
The couplings in these channels are shown in Fig.~\ref{fig:coupling}
and the bound state energies in Fig.~\ref{fig:BSeft},
each as functions of $\LambdaEFT$.
We see that the running of the coupling constant in the $^1$S$_0$ channel
is quite mild throughout the range of $\LambdaEFT$. 
In contrast, the running in each of the
other channels displays a limit-cycle pattern.
With each cycle to $+\infty$ a new bound state emerges, at energies
bound by two hundred MeV or much more, which place them outside the 
low-energy domain of the EFT.

\begin{figure*}[tbh!]
  \includegraphics[width=\movieplotwidth]{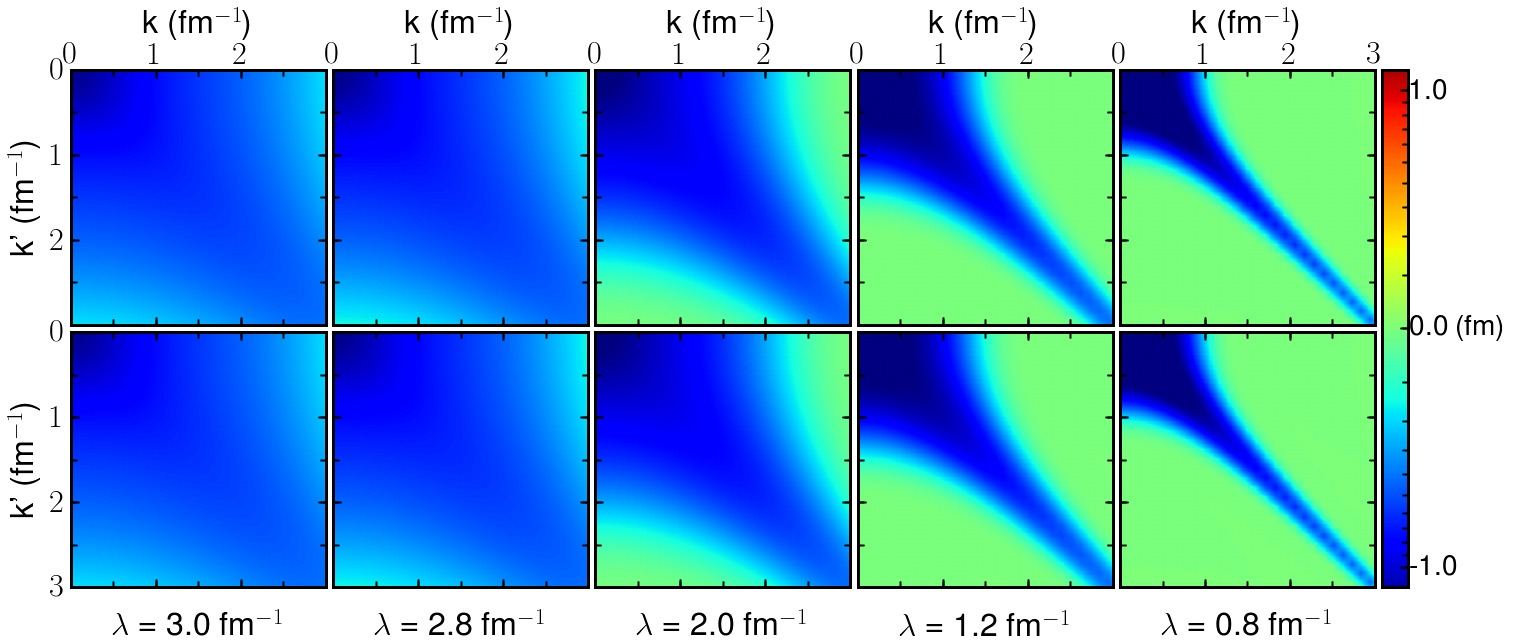}
  \vspace*{-.10in}
  \caption{(color online) Contour plot of $V_{\lambda}(k,k^\prime)$ in the $^1$S$_0$
channel for $\chi_{\rm EFT}$ with a cutoff of $\LambdaEFT =
8.0\,{\rm fm}^{-1}$ for \Trel\ (top) and \Hdiag\ (bottom) SRG
evolution.  
\label{fig:1S0Movie08}}
\end{figure*}

In Figs.~\ref{fig:1S0EFTPhaseShift}, \ref{fig:3S1EFTPhaseShift}, and \ref{fig:3P0EFTPhaseShift}
we plot the phase shifts for representative values of $\LambdaEFT$
up to a lab energy of 100\,MeV.
Because the EFT is only evaluated at leading order, we should not be
surprised at large deviations compared to experiment (here represented by the Nijmegen partial wave analysis).  However, our interest 
is not in establishing how well experiment is reproduced, but in how close the predictions of the various Hamiltonians are to each other
as a function of interaction energy.
This will be relevant when we examine universality below.


\section{Results}\label{sec:results}

In this section, we evolve LO chiral EFT potentials at representative
cutoffs using flow equations
with generators based on both $\Trel$ and $\Hdiag$.
We start with the $^1$S$_0$ channel, for which there are no spurious
bound states.  As illustrated in Fig.~\ref{fig:1S0Movie08} for
a cutoff of $\LambdaEFT = 8.0\,{\rm fm}^{-1}$, the evolved potentials for  \Trel\ and \Hdiag\ are practically indistinguishable.  This is true at any cutoff for this channel. 
More generally we have found it to be true for any channel for which 
there are no spurious bound states.

\begin{figure*}[ptbh!]
  \includegraphics[width=\movieplotwidth]{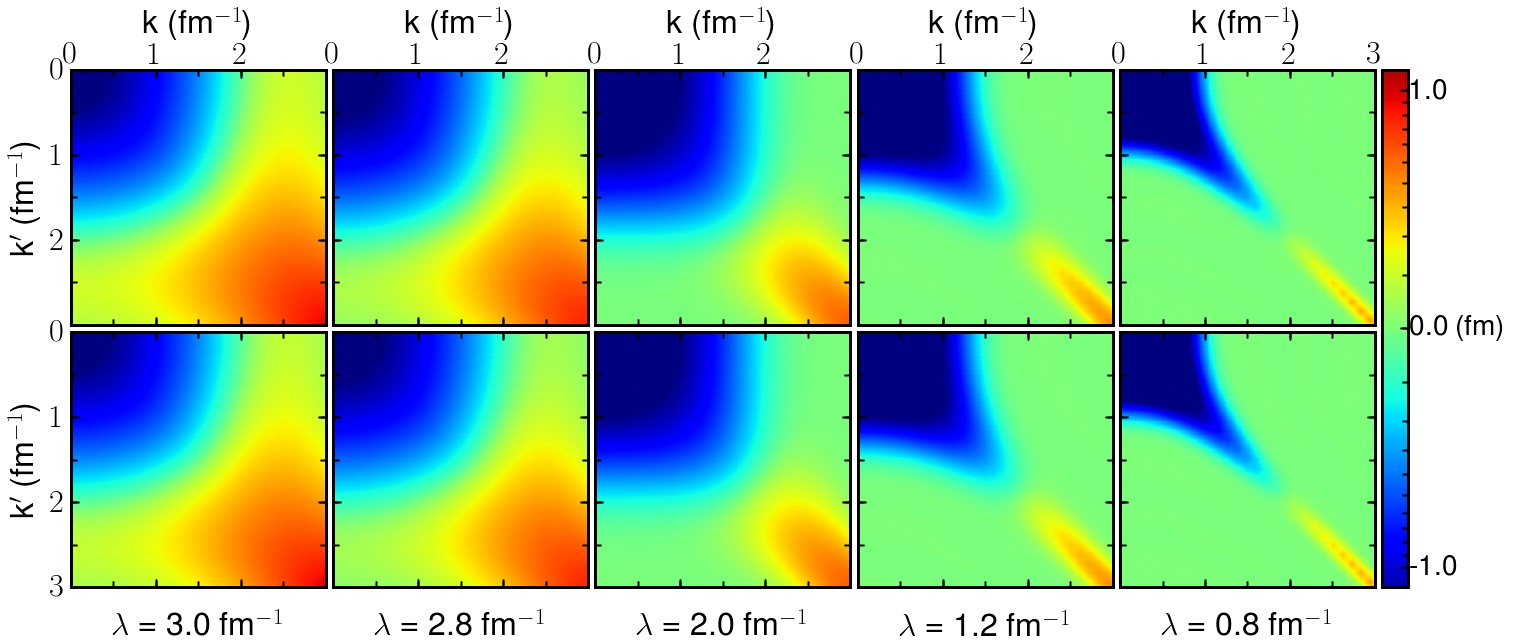}
  \vspace*{-.10in}
  \caption{(color online) Contour plot of $V_{\lambda}(k,k^\prime)$ in the $^3$S$_1$
channel for $\chi_{\rm EFT}$ with a cutoff of $\LambdaEFT =
4.0\,{\rm fm}^{-1}$ for \Trel\ (top) and \Hdiag\ (bottom) SRG
evolution.
\label{fig:3S1Movie04}}
\vspace*{0.1in}
  \includegraphics[width=\movieplotwidth]{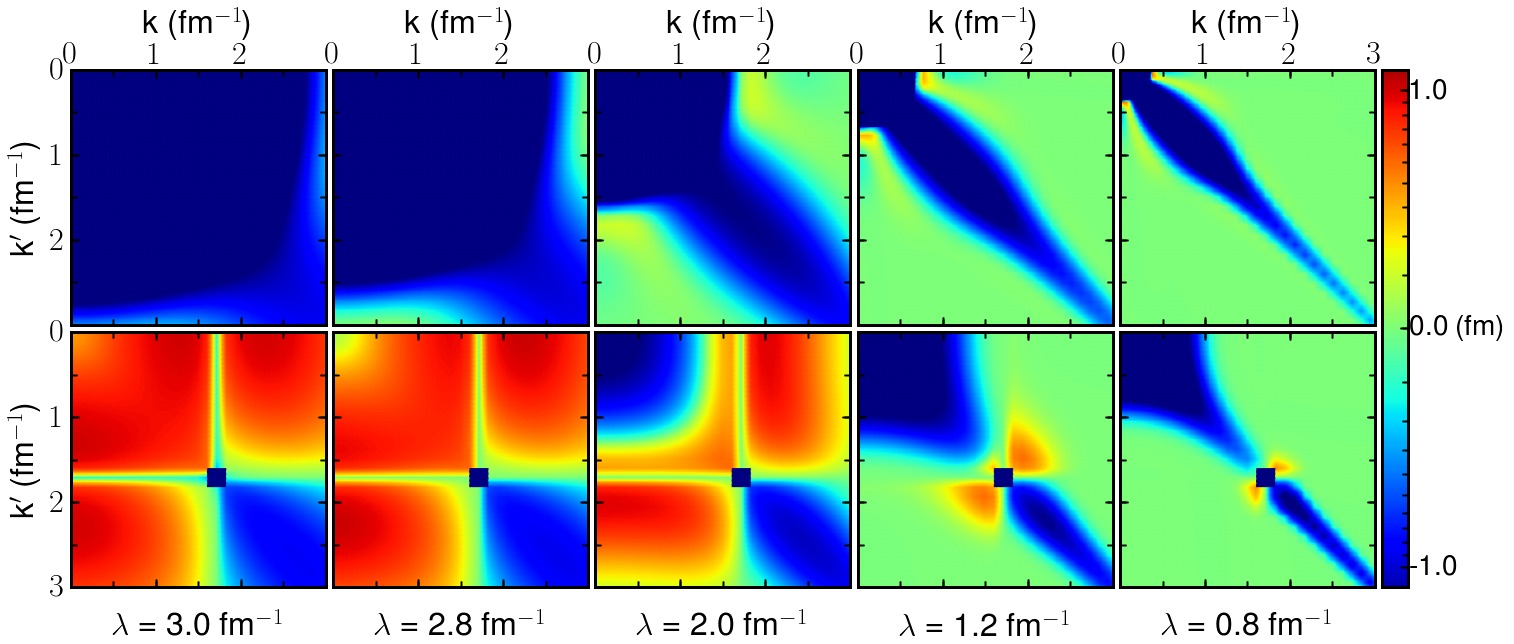}
  \vspace*{-.10in}
  \caption{(color online) Contour plot of $V_{\lambda}(k,k^\prime)$ in the $^3$S$_1$
channel for $\chi_{\rm EFT}$ with a cutoff of $\LambdaEFT =
9.0\,{\rm fm}^{-1}$ for \Trel\ (top) and \Hdiag\ (bottom) SRG
evolution.  
  \label{fig:3S1Movie09}}
\vspace*{0.1in}
  \includegraphics[width=\movieplotwidth]{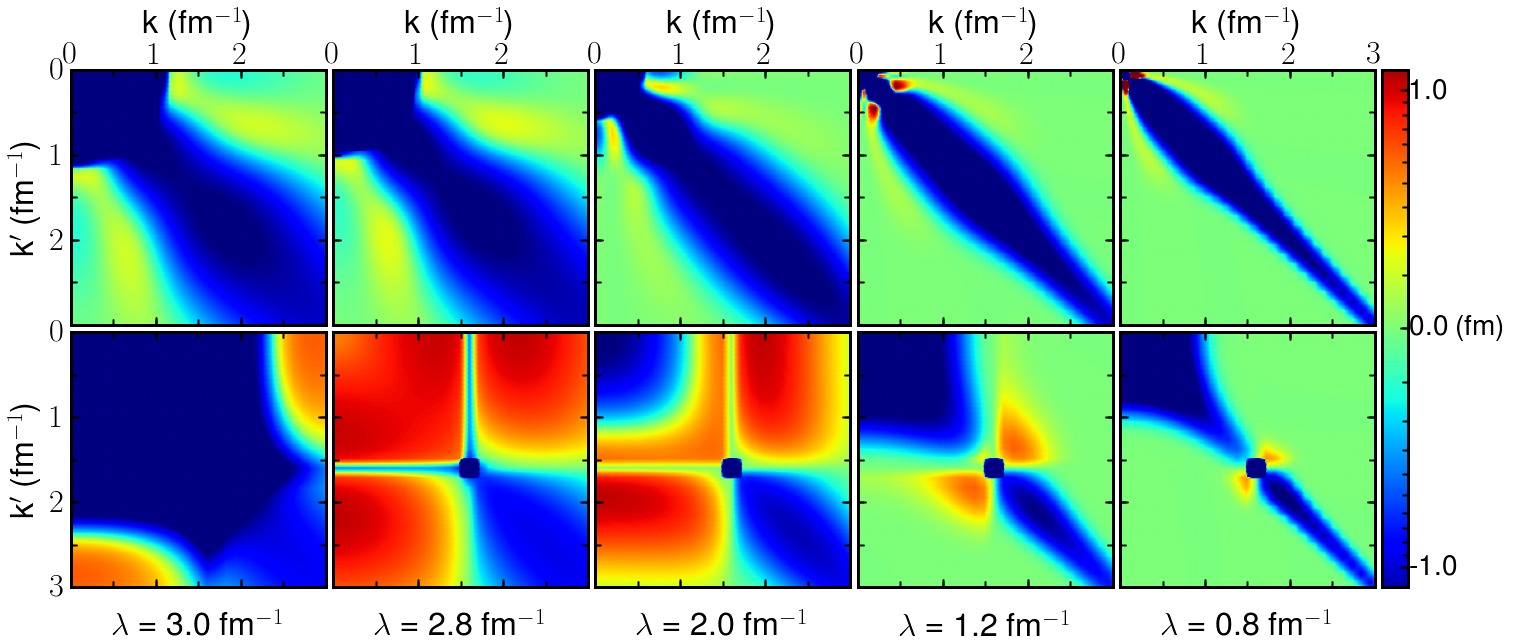}
  \vspace*{-.10in}
  \caption{(color online) Contour plot of $V_{\lambda}(k,k^\prime)$ in the $^3$S$_1$
channel for $\chi_{\rm EFT}$ with a cutoff of $\LambdaEFT =
20.0\,{\rm fm}^{-1}$ for \Trel\ (top) and \Hdiag\ (bottom) SRG
evolution.
  \label{fig:3S1Movie20}}
\end{figure*}

A very different story as a function of $\LambdaEFT$
is found in the coupled $^3$S$_1$--$^3$D$_1$
channel, as seen in Figs.~\ref{fig:3S1Movie04}, \ref{fig:3S1Movie09},
and \ref{fig:3S1Movie20}, which compare the \Trel\ and \Hdiag\
evolutions for $\LambdaEFT = 4.0$, $9.0$, and $20.0\,\fmi$, respectively.
The initial $\LambdaEFT = 4.0\,\fmi$
interaction supports only the deuteron bound state.  Without any
spurious states to decouple, the evolutions generated by \Trel\ and
\Hdiag\ are nearly identical to each other, as with $^1$S$_0$.
However, in Fig.~\ref{fig:3S1Movie09} there is now a spurious
state to decouple and so the SRG evolution for \Trel\ and \Hdiag\ are
substantially different.  

Both the \Trel\ and \Hdiag\ evolutions steadily drive the
matrix toward a diagonal form.  In the process, the low momentum part of the matrix is driven towards large negative values.  With \Trel\ this is unabated, while in the \Hdiag\ evolution it reverses momentarily and deposits an isolated negative eigenvalue on the diagonal, which corresponds to the complete decoupling of this spurious state~\cite{Glazek:2008pg}.
The low-momentum potential in the last two panels of Fig.~\ref{fig:3S1Movie09} for \Trel\ (bottom) is 
qualitatively different from the corresponding panels for 
$\LambdaEFT = 4.0\,\fmi$ in Fig.~\ref{fig:3S1Movie04} while the
corresponding \Hdiag\ panels are similar for momenta below the
spurious state matrix element.

Figure.~\ref{fig:3S1Movie20} shows what happens when a second spurious
state is supported after increasing the EFT cutoff to  $\LambdaEFT =
20.0\,{\rm fm}^{-1}$. 
Again \Trel\ and \Hdiag\ differ significantly but the latter is
similar to \Hdiag\ in the other figures.  If we expanded the momentum
scales, we would find that the deepest state is first decoupled, with
the diagonal element found at much larger $k$,  and then the shallower
spurious state is decoupled at smaller $\lambda$.

\begin{figure*}[ptbh!]
  \includegraphics[width=\movieplotwidth]{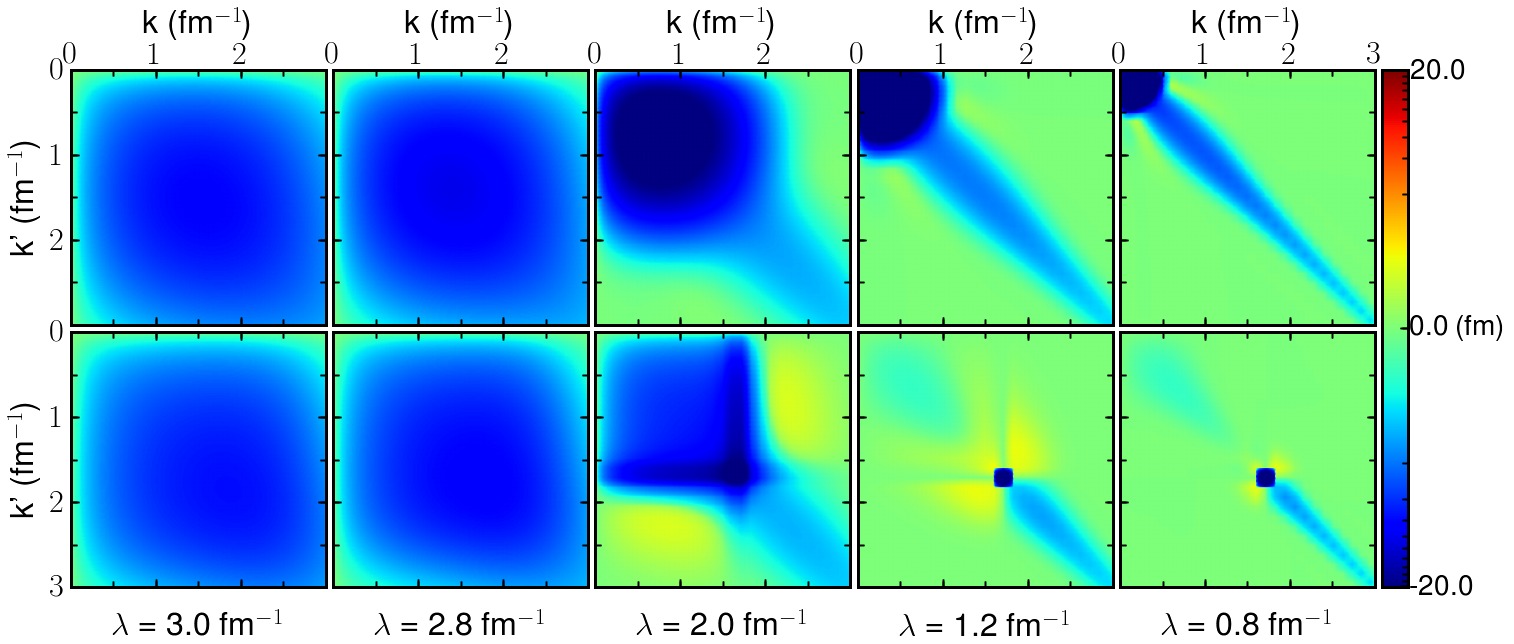}
  \vspace*{-.10in}
  \caption{(color online) Contour plot of $V_{\lambda}(k,k^\prime)$ in the $^3$P$_0$
channel for $\chi_{\rm EFT}$ with a cutoff of $\LambdaEFT =
4.0\,{\rm fm}^{-1}$ for \Trel\ (top) and \Hdiag\ (bottom) SRG
evolution.  
\label{fig:3P0Movie04}}
\vspace*{0.1in}
  \includegraphics[width=\movieplotwidth]{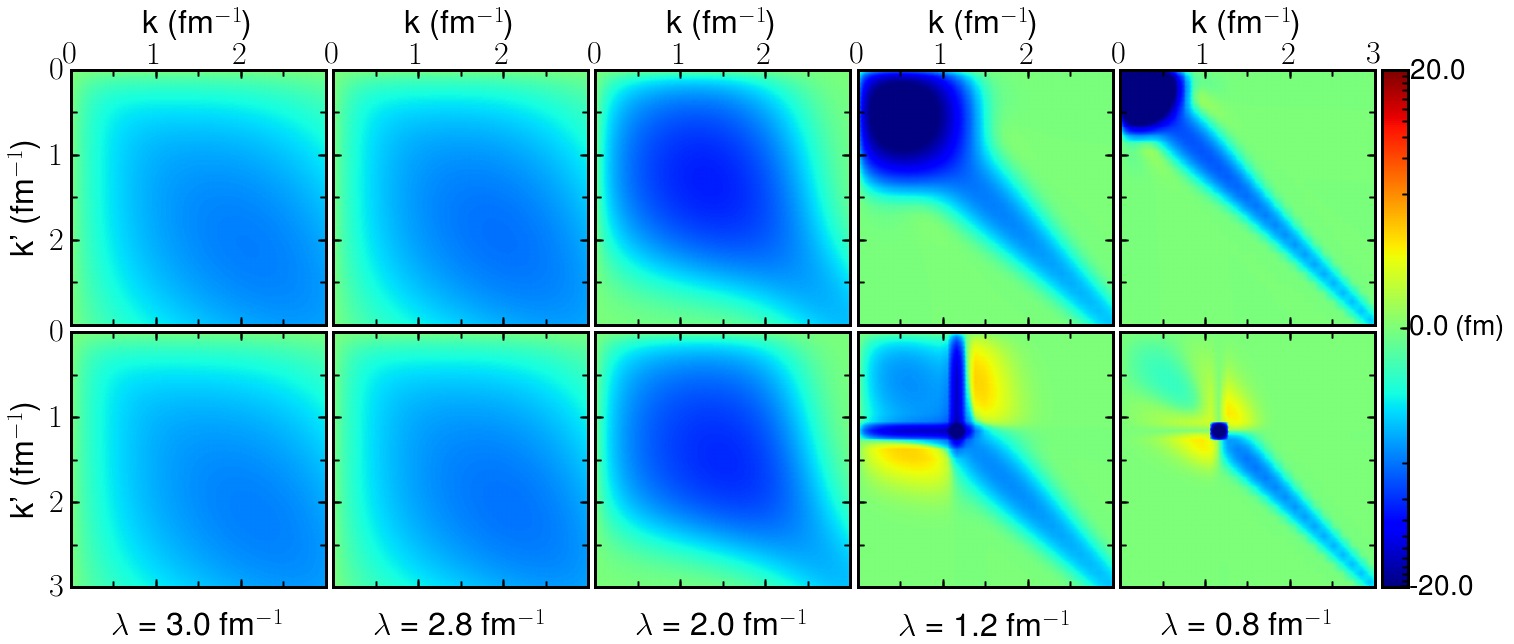}
  \vspace*{-.10in}
  \caption{(color online) Contour plot of $V_{\lambda}(k,k^\prime)$ in the $^3$P$_0$
channel for $\chi_{\rm EFT}$ with a cutoff of $\LambdaEFT =
8.0\,{\rm fm}^{-1}$ for \Trel\ (top) and \Hdiag\ (bottom) SRG
evolution.
  \label{fig:3P0Movie09}}
\vspace*{0.1in}
  \includegraphics[width=\movieplotwidth]{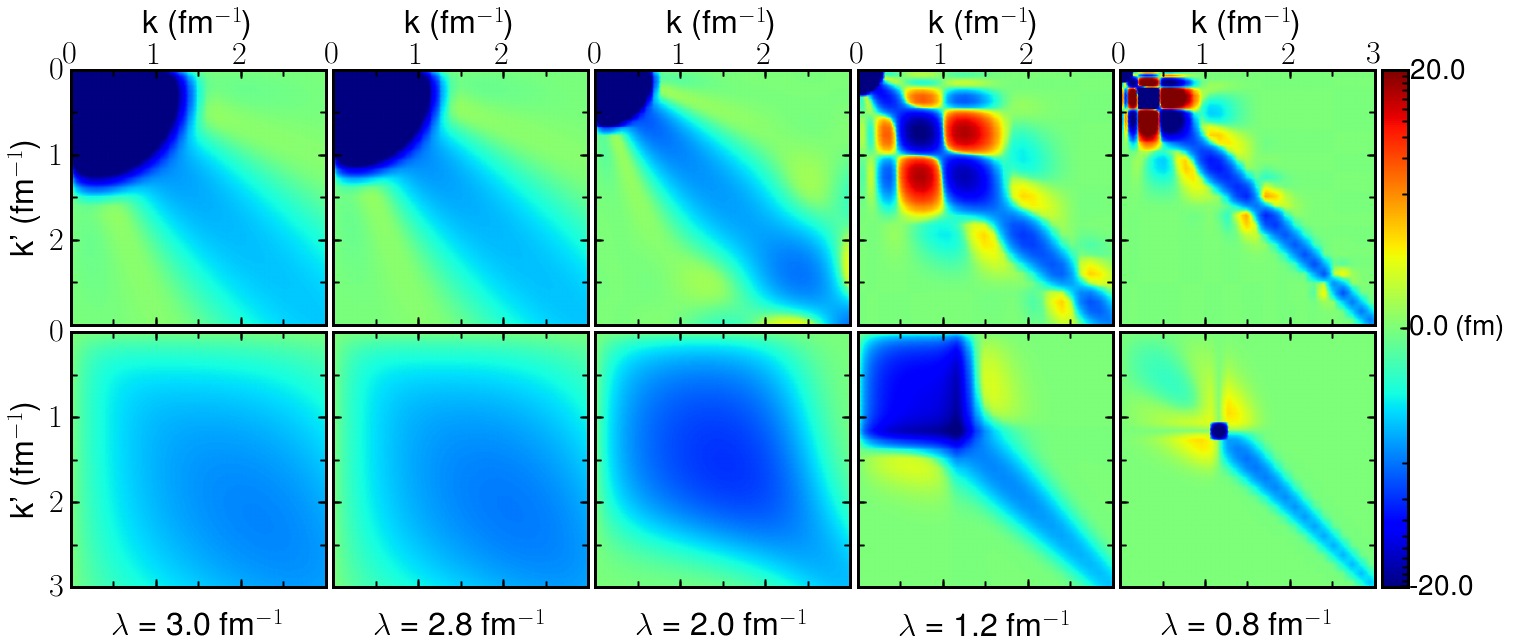}
  \vspace*{-.10in}
  \caption{(color online) Contour plot of $V_{\lambda}(k,k^\prime)$ in the $^3$P$_0$
channel for $\chi_{\rm EFT}$ with a cutoff of $\LambdaEFT =
20.0\,{\rm fm}^{-1}$ for \Trel\ (top) and \Hdiag\ (bottom) SRG
evolution.
\label{fig:3P0Movie20}}
\end{figure*}

For the uncoupled $^3$P$_0$ channel, the tensor is also attractive and
increasing the cutoff of the EFT leads to a spurious bound state for
$\LambdaEFT$ between 3 and $4\,\fmi$.  
In Figs.~\ref{fig:3P0Movie04}, \ref{fig:3P0Movie09}, and
\ref{fig:3P0Movie20} we trace the evolution for two cutoffs above the
threshold for a single bound state ($\LambdaEFT = 4\,\fmi$ and
$\LambdaEFT = 8\,\fmi$), and for a third above the threshold for two
bound states ($\LambdaEFT = 20\,\fmi$).
Results for low cutoffs with no bound states are not shown, but have
the same pattern as for $^1$S$_0$.
With a spurious state to decouple, the SRG evolutions for \Trel\ and
\Hdiag\ at the lowest  $\LambdaEFT$ are again substantially different,
just as for the $^3$S$_1$--$^3$D$_1$ channel.
We see particularly non-universal evolution of the low-momentum part
of the \Trel\ potentials while the \Hdiag\ potentials are similar (more quantitative comparisons are made
below).  Note that the diagonalization of the spurious bound
states do not occur at the same diagonal momentum.

\begin{figure*}[ptbh!]
 \includegraphics[width=\universalityplotwidth]{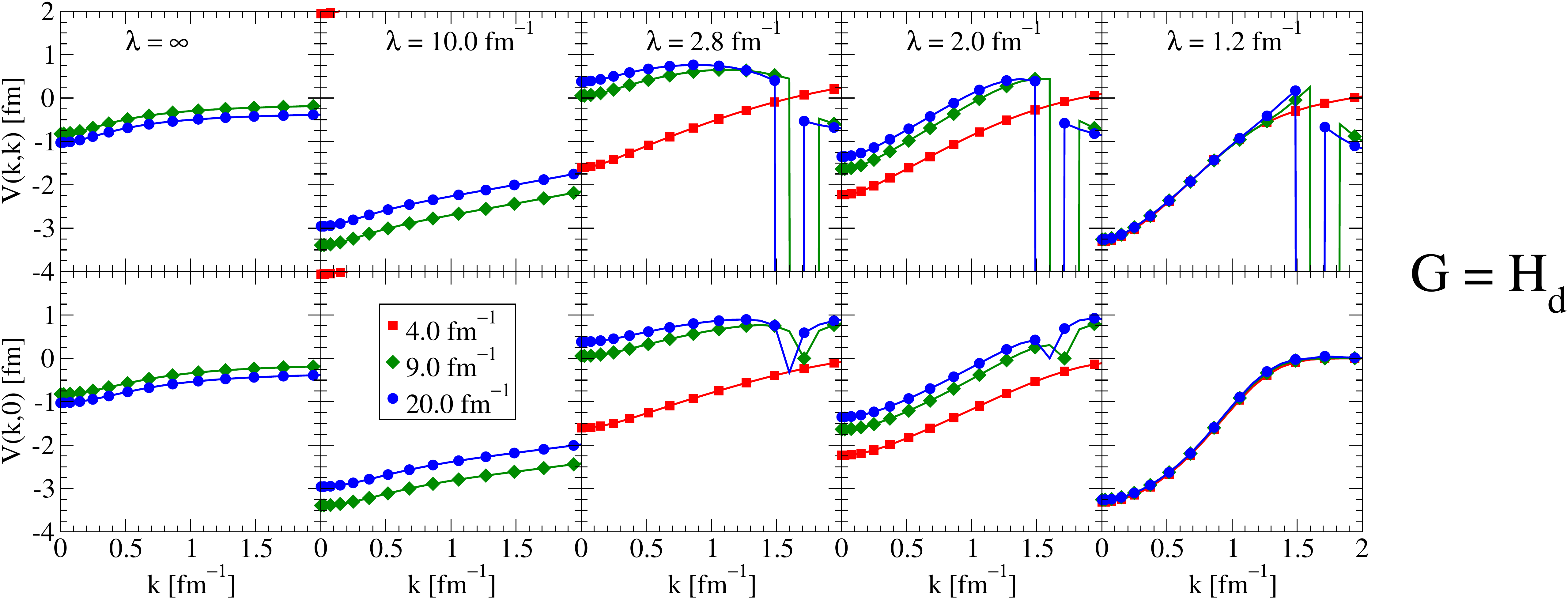}
  \vspace*{-.10in}
  \caption{(color online) The diagonal matrix elements $V_s(k,k)$ and
  fully off-diagonal matrix elements $V_s(k,0)$  for the $^3{\rm S}_1$
  channel are plotted against momentum $k$ after evolving using the
  Wegner generator to a range of final flow parameters
  $\lambda = s^{-1/4}$.  For each $\lambda$, results for five cutoff
  values $\LambdaEFT$ are shown (in some cases they are beyond the
  limits of the plot).
  \label{fig:EVOL_HD}}
\end{figure*}

\begin{figure*}[ptbh!]
  \includegraphics[width=\universalityplotwidth]{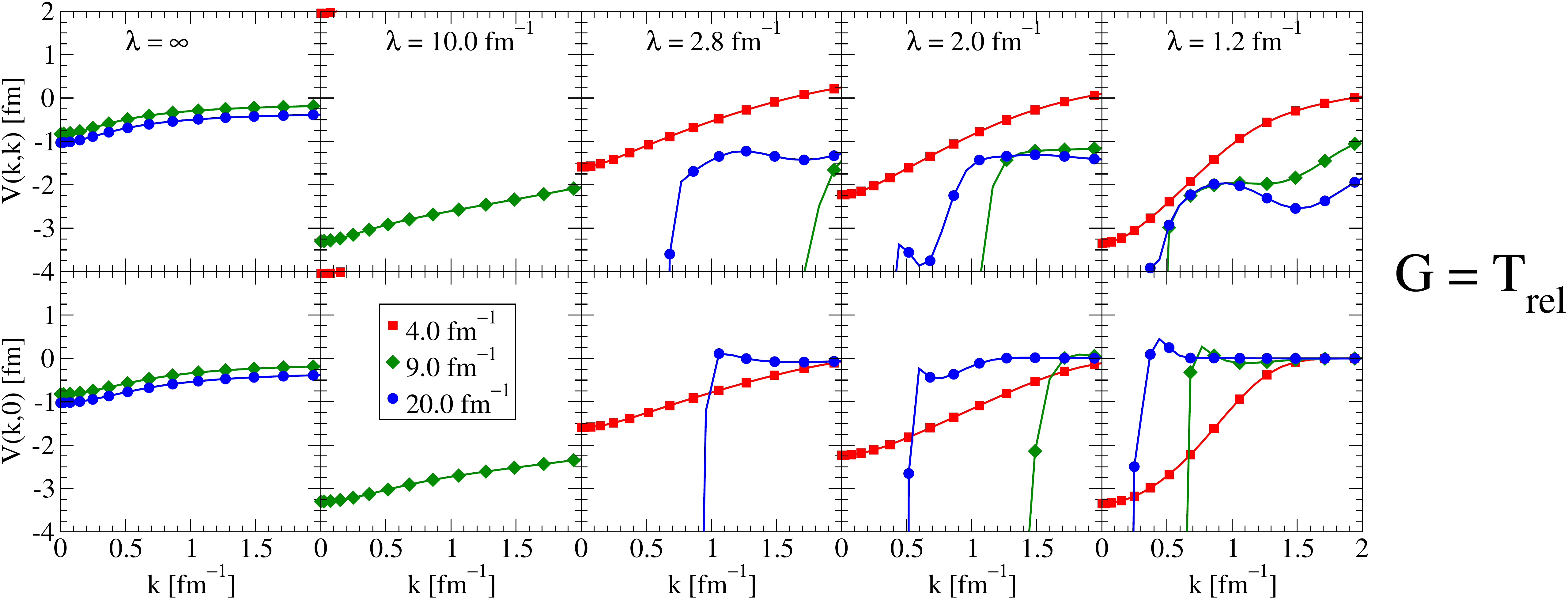}
  \vspace*{-.10in}
  \caption{(color online) The diagonal matrix elements $V_s(k,k)$
  and    fully off-diagonal matrix elements $V_s(k,0)$   for the
  $^3{\rm S}_1$ channel are plotted against momentum $k$ after
  evolving using the \Trel--based generator to a range of final flow
  parameters $\lambda = s^{-1/4}$.  For each $\lambda$, results for
  five cutoff values $\LambdaEFT$ are shown (in some cases they are
  beyond the limits of the plot).
  \label{fig:EVOL_TREL}}
\end{figure*}

The quantitative evolution can be characterized by focusing on the
diagonal and fully off-diagonal matrix elements separately. In
Figs.~\ref{fig:EVOL_HD} and \ref{fig:EVOL_TREL} the diagonal and fully
off-diagonal matrix elements of the interactions in the $^3$S$_1$
are plotted for
\Hdiag\ and \Trel, respectively, as representative examples.
The decoupling of high and low momentum degrees of freedom is a
prevalent feature in the Wegner evolution whether there is
a deep bound state or not.  
The result of this decoupling is that the low momentum portion of the
matrix becomes universal, with a collapse of the original potentials
to almost the same low-$k$ dependence for sufficiently low
$\lambda$.  
In contrast, with the $\Trel$~evolution the low-momentum
diagonal matrix elements are always quite distinct. 
The plots showing the off-diagonal edge manifest the distinction
most clearly
because the non-universal
off-diagonal elements above $\lambda$ are driven to zero.
Similar behavior is observed in other channels.
The degree of collapse with the Wegner evolution is correlated
with the level of agreement of the phase shifts in that channel.

\begin{figure*}[tbh!]

\subfigure[]{\includegraphics[width=0.9\columnwidth]{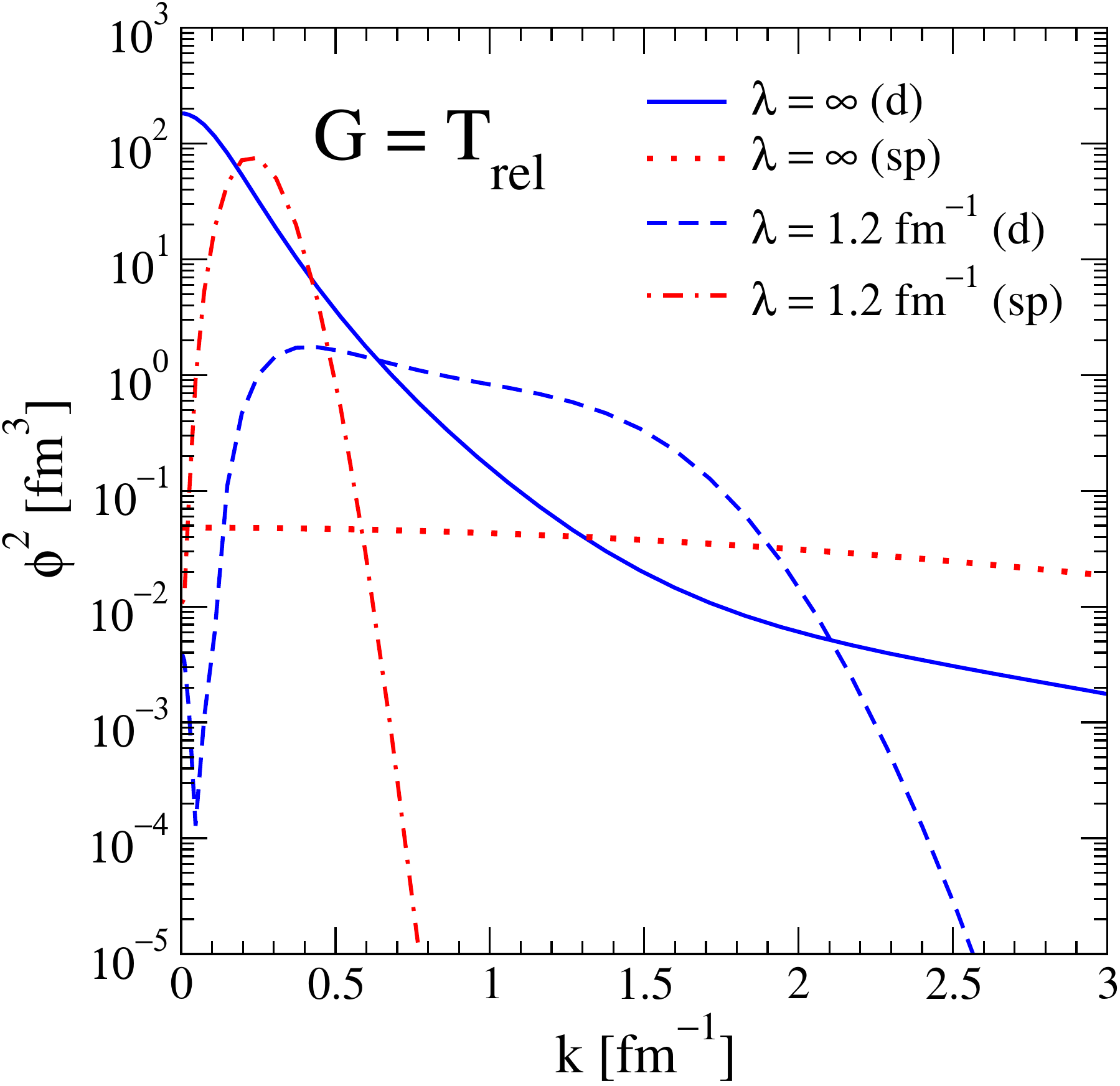}}\hspace*{.2in}
\subfigure[]{\includegraphics[width=0.9\columnwidth]{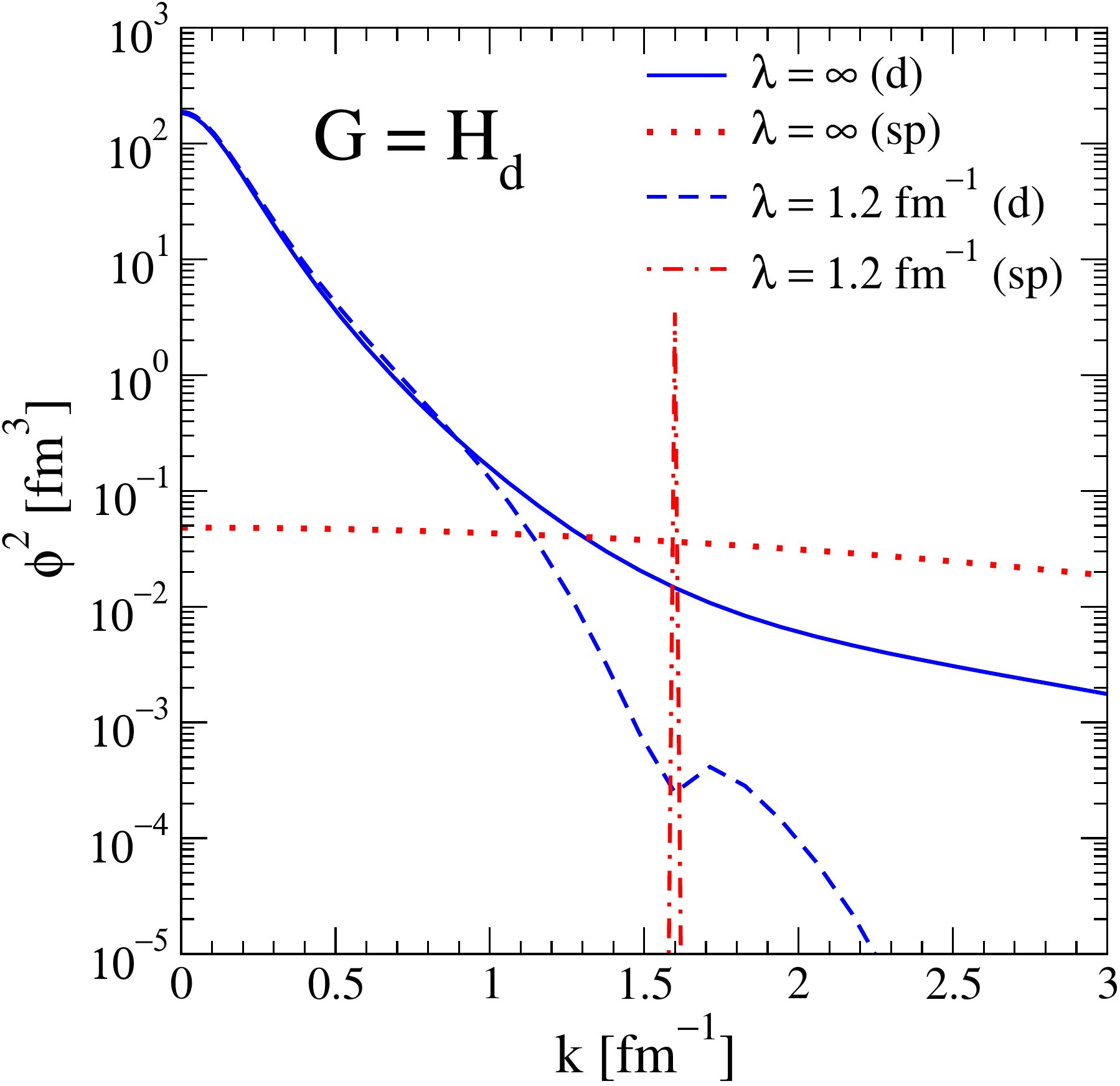}} 
  \vspace*{-.10in}
  \caption{(color online) Momentum probability densities in the
  coupled $^3$S$_1$--$^3$D$_1$ channel with $\LambdaEFT = 20\,\fmi$
  for the deuteron (d) and shallowest spurious state (sp).
  On the left are the initial and evolved densities using the \Trel\
  generator evolved to $\lambda = 1.2\,\fmi$ while on the right are
  the same densities but evolved using the \Hdiag\ generator.
  \label{fig:momdists} } 
\end{figure*}

Figure~\ref{fig:momdists} shows similarly striking consequences of the
choice of generator for the momentum distributions of the bound states
in the $^3$S$_1$--$^3$D$_1$ channel, with $\LambdaEFT = 20\,\fmi$ and
an evolution  to $\lambda = 1.2\,\fmi$ chosen as a  representative
case.
The initial deuteron momentum distribution has a nearly exponential
tail while the spurious state is essentially flat in the region of
momenta plotted (reflecting the confinement of the coordinate-space
wavefunction).
Their fates after evolution with \Trel\ and \Hdiag\ are starkly
different.
In the former case (left panel), the spurious state has had its
strength pushed to low momentum while the deuteron becomes broadly
distributed and qualitatively changed at low momentum.  In the latter
case (right panel), the spurious state is confined to a small region
in momentum near $k=1.6\,\fmi$ while the deuteron momentum distribution at low momentum is essentially unchanged.
Similar results are found with other choices of $\LambdaEFT$.

\begin{figure*}[ph!]
  \includegraphics[width=\phaseplotwidth]{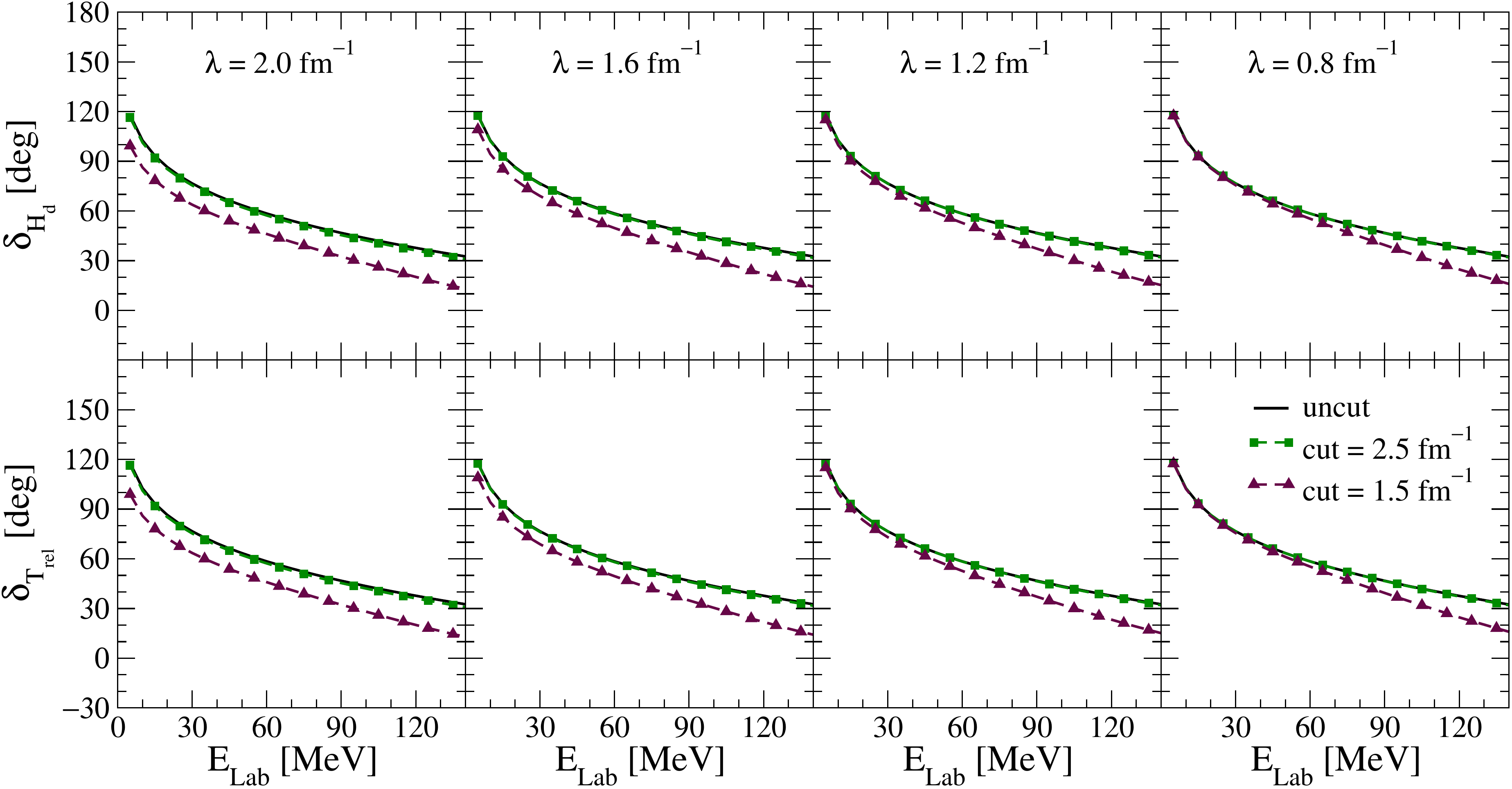}
  \vspace*{-.10in}
  \caption{(color online) Phase shifts in the coupled $^3$S$_1$--$^3$D$_1$ channel 
  using the $\LambdaEFT = 4.0\,\fmi$ potential, testing
  \Hdiag\ (top) and \Trel\ (bottom) SRG decoupling by first cutting off the potential.
  \label{fig:phases3S14p0}}
\vspace*{0.1in}  
%
  \includegraphics[width=\phaseplotwidth]{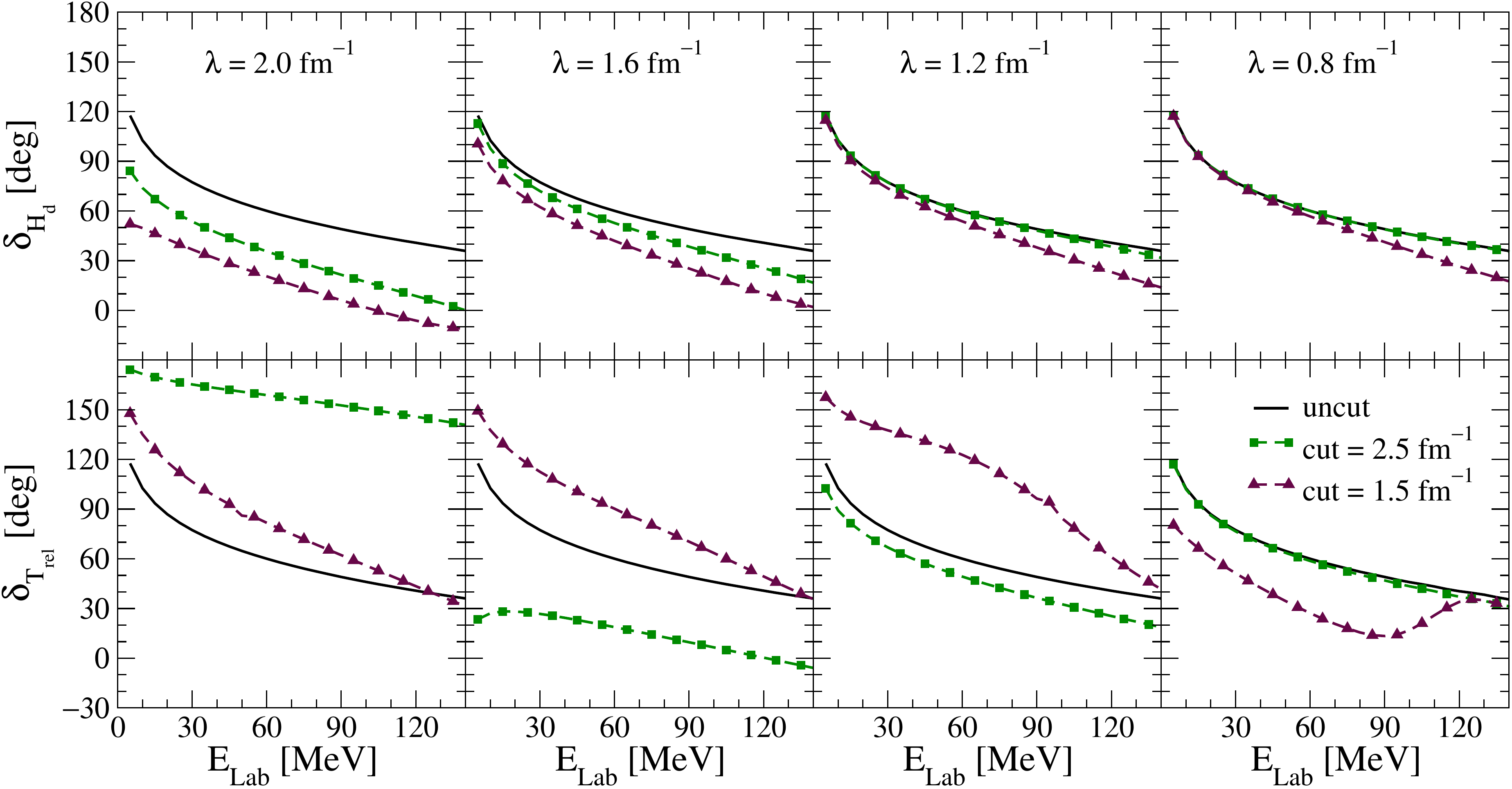}
  \vspace*{-.10in}
  \caption{(color online) Phase shifts in the coupled $^3$S$_1$--$^3$D$_1$ channel 
  using the $\LambdaEFT = 9.0\,\fmi$ potential, testing
  \Hdiag\ (top) and \Trel\ (bottom) SRG decoupling by first cutting off the potential.
  \label{fig:phases3S19p0}}
\vspace*{0.1in}  
%
  \includegraphics[width=\phaseplotwidth]{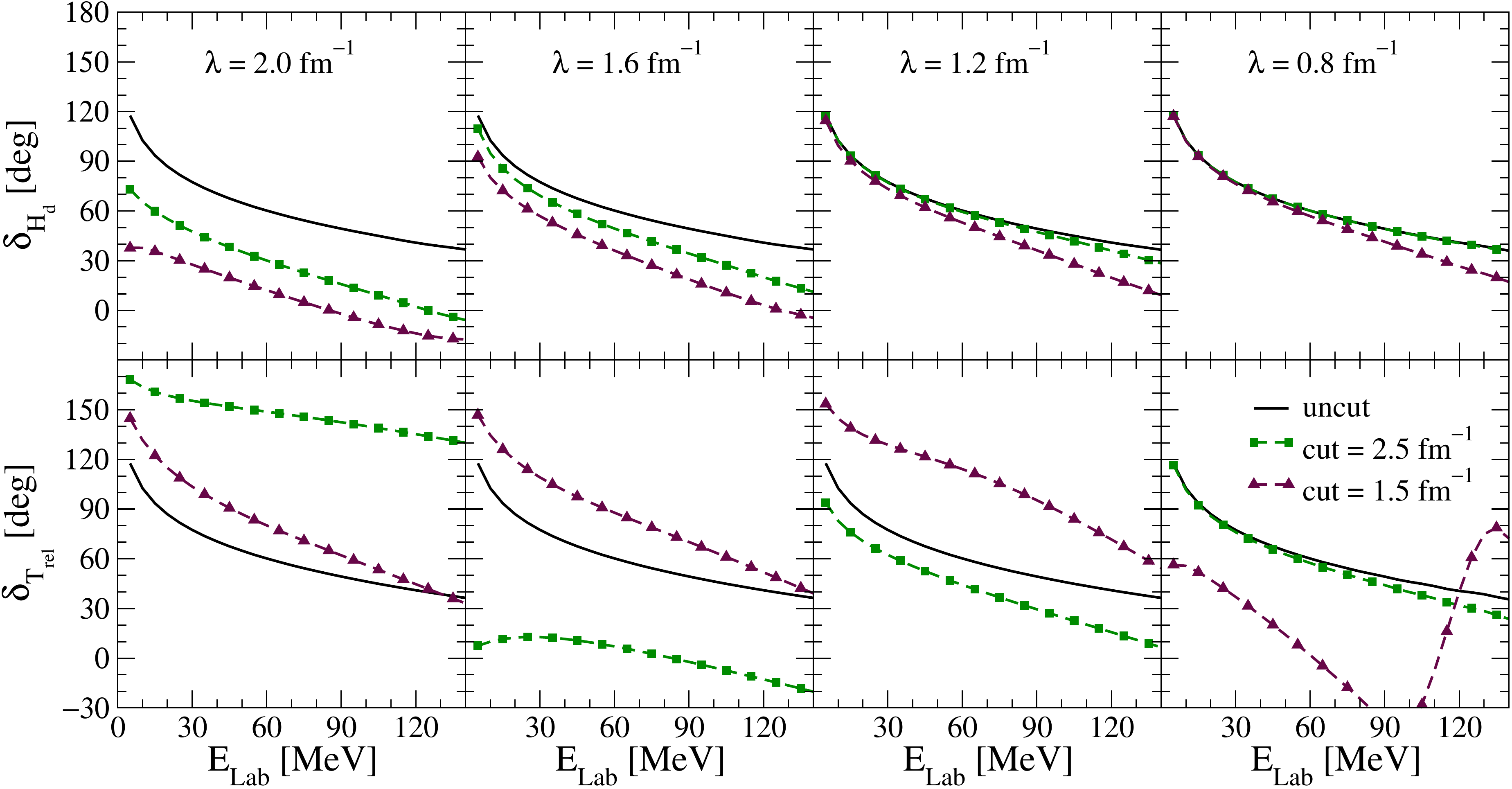}
  \vspace*{-.10in}
  \caption{(color online) Phase shifts in the coupled $^3$S$_1$--$^3$D$_1$ channel 
  using the $\LambdaEFT = 20.0\,\fmi$ potential, testing
  \Hdiag\ (top) and \Trel\ (bottom) SRG decoupling by first cutting off the potential.
  \label{fig:phases3S120p0}}
\end{figure*}

The phase shifts for a given $\LambdaEFT$ are unchanged as a result of
a unitary SRG evolution.  However, we can test the degree of
decoupling in different cases by cutting the evolved potential (i.e.,
setting it to zero for all $k,k'$ above a cut momentum) and looking at
the degradation in the phase shifts.
This is shown in Figs.~\ref{fig:phases3S14p0},
\ref{fig:phases3S19p0},  \ref{fig:phases3S120p0} for the $^3$S$_1$
phases shifts evolved from $\LambdaEFT = 4$, 9, and $20\,\fmi$,
respectively.
The cut is implemented by multiplying each matrix element by the factor $e^{-(k/\Lambda)^{2m}} e^{-(k^\prime/\Lambda)^{2m}}$.  
Results for $m=2,\,4,\,8$ are similar; we show $m=4$ only.
In general, successful decoupling means that the phase shifts 
calculated with a potential evolved to $\lambda$ are
unchanged at low energy if $\lambda < \Lambda$
(of course they will deviate for energies above the cut). 

If there are no spurious bound states, decoupling works for either
\Trel\ or \Hdiag\ evolution, as seen in Fig.~\ref{fig:phases3S14p0}.
In particular, the phase shifts are essentially unchanged for $\lambda
\leq 2.0\,\fmi$ after a $2.5\,\fmi$ cut and agree at low energy after
a $1.5\,\fmi$ cut for sufficiently low $\lambda$. The pattern of
decoupling is similar for \Hdiag-evolved potentials when there are
spurious bound states, except that the onset of decoupling is delayed
until $\lambda$ is sufficiently below where the bound state is
deposited on the diagonal.  Thus, in the top panels of
Figs.~\ref{fig:phases3S19p0} and \ref{fig:phases3S120p0}, decoupling
is not achieved until $\lambda = 1.2\,\fmi$, compared to $\lambda =
2.0\,\fmi$ in Fig.~\ref{fig:phases3S14p0} for the $2.5\,\fmi$ cut.  This may slow convergence
in few- or many-body calculations; this remains to be investigated. The decoupling patterns for phases shifts calculated from \Trel-evolved potentials with spurious bound states show serious distortions at the lower cut.
sufficiently low the phase shifts are again reproduced at low energy.

\section{Discussion}\label{sec:discussion}

We have documented
several different outcomes from evolution with generators using
$H_d$ and $\Trel$.
For $\Trel$, the different classes of low-momentum Hamiltonians depend
on the number of spurious states, while for $\Hdiag$ there is a single
universal class for the low-energy part, after sufficient evolution.
There is corruption of the low-momentum probability distribution for
$\Trel$ and the physical low-lying bound state is still coupled and
non-universal.  
In contrast,
with \Hdiag\ the spurious state completely decouples and the deuteron
wave function is unchanged at low momentum.
Note that despite these unhappy features for $\Trel$, the results
for all observables
are formally the same because we still have a unitary transformation acting on a renormalized Hamiltonian.
In practice, however, numerical computation may lose all precision for observables influenced by the near-diagonalized part and our ability
to truncate a basis expansion will be compromised.

Although it is clear that decoupling increases with decreasing $\lambda$ when using
\Hdiag,  it may not be so useful in practice.
The binding energies of the spurious states seen here are very large, yet the decoupling momenta for the least bound ones are still comparable or smaller than values where SRG evolution is usually halted because of concerns about the growth of many-body forces. 
Will the remaining distortions spoil the advantages?
For example, are variational calculations still effective and is convergence
in basis expansions still accelerated?
Answering these questions will require extending our tests to
$A \geq 3$.
Also, the question of where a bound state is placed on the diagonal
in momentum representation is still open.

In Sect.~\ref{sec:background} we invoked the
claim that the deep bound states are outside the range of the EFT and as a result should not disturb the low-energy EFT predictions.  
Our results provide an explicit test of this principle.
In particular, the SRG unitary flow does not change the S-matrix 
and yet with an \Hdiag-based generator we are able to create a decoupled version of the Hamiltonian.  Because it 
becomes decoupled at a finite momentum scale $\lambda$ and has the same universal form in the low-energy region as potentials with no spurious bound states, we have achieved the desired demonstration.

An important open question is the impact on SRG evolution of bound
states for $A > 2$.  For example, at what $\lambda$ do we expect to
have problems evolving with a \Trel-based generator for the triton,
$^4$He, or nuclear matter? The choice between \Trel\ and \Hdiag\ for
building the SRG generator is just one of many possible useful choices
to control the flow of the Hamiltonian.   What other generators
achieve what \Hdiag\ does but allow more freedom?  Work is in
progress to address these issues.
  
\section{Summary}\label{sec:summary}

In this paper, we have applied SRG evolution flow equations that have
been uniformly successful with commonly used initial NN interactions
to a new class of initial potentials. These potentials are from
renormalized leading-order chiral EFT  and have much higher
momentum-space cutoffs which, combined with singular pion exchange,
lead to spurious bound states emerging in channels where there is a
non-zero, attractive tensor force. We have tested flow properties such
as decoupling and the approach to universal form.

While the choice of $\Trel$ in the SRG generator has been successful
for nuclear applications to date, this has been the fortuitous result
of shallow bound states that are not disturbed for the $\lambda$'s
considered. The presence of deep bound states introduces problems
first emphasized by Glazek and Perry~\cite{Glazek:2008pg}. The generalization to \Hdiag\
restores the good features such as decoupling and flow to a
universal form as $\lambda$ decreases.

An important moral is that even a comparatively subtle change in the
generator can have a substantial effect on the flow of  the
Hamiltonian.  While in principle observables are always unchanged
because we are making unitary transformations, in practice there are
always truncations that make the transformations only approximately
unitary.  It is critical to minimize the impact of these
approximations as we extend calculations to more particles.  At the same
time, the degree of perturbativeness and convergence properties of the
Hamiltonian can also be significantly affected.  Future work includes
looking at few-body systems and using the SRG in further
investigations of EFT. 


\vspace*{.1in}

\begin{acknowledgments}
We thank E.~Anderson, S.~Bogner, E.~Jurgenson, W.~Li, and A.~Schwenk
for useful comments and discussions.
This work was supported in part by the National Science Foundation
under Grant Nos.~PHY--0653312 and PHY--1002478, 
the UNEDF SciDAC Collaboration under DOE Grant 
DE-FC02-07ER41457, 
and in part by an award from the Department of Energy (DOE) Office of Science Graduate Fellowship Program (DOE SCGF). The DOE SCGF Program was made possible in part by the American Recovery and Reinvestment Act of 2009.  The DOE SCGF program is administered by the Oak Ridge Institute for Science and Education for the DOE. ORISE is managed by Oak Ridge Associated Universities (ORAU) under DOE contract number DE-AC05-06OR23100.
\end{acknowledgments}

\bibliography{vlowk_refs}

\begin{thebibliography}{22}%
\makeatletter
\providecommand \@ifxundefined [1]{%
 \@ifx{#1\undefined}
}%
\providecommand \@ifnum [1]{%
 \ifnum #1\expandafter \@firstoftwo
 \else \expandafter \@secondoftwo
 \fi
}%
\providecommand \@ifx [1]{%
 \ifx #1\expandafter \@firstoftwo
 \else \expandafter \@secondoftwo
 \fi
}%
\providecommand \natexlab [1]{#1}%
\providecommand \enquote  [1]{``#1''}%
\providecommand \bibnamefont  [1]{#1}%
\providecommand \bibfnamefont [1]{#1}%
\providecommand \citenamefont [1]{#1}%
\providecommand \href@noop [0]{\@secondoftwo}%
\providecommand \href [0]{\begingroup \@sanitize@url \@href}%
\providecommand \@href[1]{\@@startlink{#1}\@@href}%
\providecommand \@@href[1]{\endgroup#1\@@endlink}%
\providecommand \@sanitize@url [0]{\catcode `\\12\catcode `\$12\catcode
  `\&12\catcode `\#12\catcode `\^12\catcode `\_12\catcode `\%12\relax}%
\providecommand \@@startlink[1]{}%
\providecommand \@@endlink[0]{}%
\providecommand \url  [0]{\begingroup\@sanitize@url \@url }%
\providecommand \@url [1]{\endgroup\@href {#1}{\urlprefix }}%
\providecommand \urlprefix  [0]{URL }%
\providecommand \Eprint [0]{\href }%
\providecommand \doibase [0]{http://dx.doi.org/}%
\providecommand \selectlanguage [0]{\@gobble}%
\providecommand \bibinfo  [0]{\@secondoftwo}%
\providecommand \bibfield  [0]{\@secondoftwo}%
\providecommand \translation [1]{[#1]}%
\providecommand \BibitemOpen [0]{}%
\providecommand \bibitemStop [0]{}%
\providecommand \bibitemNoStop [0]{.\EOS\space}%
\providecommand \EOS [0]{\spacefactor3000\relax}%
\providecommand \BibitemShut  [1]{\csname bibitem#1\endcsname}%
\let\auto@bib@innerbib\@empty
\bibitem [{\citenamefont {Jurgenson}\ \emph {et~al.}(2008)\citenamefont
  {Jurgenson}, \citenamefont {Bogner}, \citenamefont {Furnstahl},\ and\
  \citenamefont {Perry}}]{Jurgenson:2007td}%
  \BibitemOpen
  \bibfield  {author} {\bibinfo {author} {\bibfnamefont {E.~D.}\ \bibnamefont
  {Jurgenson}}, \bibinfo {author} {\bibfnamefont {S.~K.}\ \bibnamefont
  {Bogner}}, \bibinfo {author} {\bibfnamefont {R.~J.}\ \bibnamefont
  {Furnstahl}}, \ and\ \bibinfo {author} {\bibfnamefont {R.~J.}\ \bibnamefont
  {Perry}},\ }\href {\doibase 10.1103/PhysRevC.78.014003} {\bibfield  {journal}
  {\bibinfo  {journal} {Phys. Rev. C}\ }\textbf {\bibinfo {volume} {78}},\
  \bibinfo {pages} {014003} (\bibinfo {year} {2008})},\ \Eprint
  {http://arxiv.org/abs/0711.4252} {arXiv:0711.4252 [nucl-th]} \BibitemShut
  {NoStop}%
\bibitem [{\citenamefont {Bogner}\ \emph {et~al.}(2010)\citenamefont {Bogner},
  \citenamefont {Furnstahl},\ and\ \citenamefont {Schwenk}}]{Bogner:2009bt}%
  \BibitemOpen
  \bibfield  {author} {\bibinfo {author} {\bibfnamefont {S.~K.}\ \bibnamefont
  {Bogner}}, \bibinfo {author} {\bibfnamefont {R.~J.}\ \bibnamefont
  {Furnstahl}}, \ and\ \bibinfo {author} {\bibfnamefont {A.}~\bibnamefont
  {Schwenk}},\ }\href {\doibase 10.1016/j.ppnp.2010.03.001} {\bibfield
  {journal} {\bibinfo  {journal} {Prog. Part. Nucl. Phys.}\ }\textbf {\bibinfo
  {volume} {65}},\ \bibinfo {pages} {94} (\bibinfo {year} {2010})},\ \Eprint
  {http://arxiv.org/abs/0912.3688} {arXiv:0912.3688 [nucl-th]} \BibitemShut
  {NoStop}%
\bibitem [{\citenamefont {Bogner}\ \emph {et~al.}(2008)\citenamefont {Bogner}
  \emph {et~al.}}]{Bogner:2007rx}%
  \BibitemOpen
  \bibfield  {author} {\bibinfo {author} {\bibfnamefont {S.~K.}\ \bibnamefont
  {Bogner}} \emph {et~al.},\ }\href {\doibase 10.1016/j.nuclphysa.2007.12.008}
  {\bibfield  {journal} {\bibinfo  {journal} {Nucl. Phys. A}\ }\textbf
  {\bibinfo {volume} {801}},\ \bibinfo {pages} {21} (\bibinfo {year} {2008})},\
  \Eprint {http://arxiv.org/abs/0708.3754} {arXiv:0708.3754 [nucl-th]}
  \BibitemShut {NoStop}%
\bibitem [{\citenamefont {Jurgenson}\ \emph {et~al.}(2009)\citenamefont
  {Jurgenson}, \citenamefont {Navratil},\ and\ \citenamefont
  {Furnstahl}}]{Jurgenson:2009qs}%
  \BibitemOpen
  \bibfield  {author} {\bibinfo {author} {\bibfnamefont {E.~D.}\ \bibnamefont
  {Jurgenson}}, \bibinfo {author} {\bibfnamefont {P.}~\bibnamefont {Navratil}},
  \ and\ \bibinfo {author} {\bibfnamefont {R.~J.}\ \bibnamefont {Furnstahl}},\
  }\href {\doibase 10.1103/PhysRevLett.103.082501} {\bibfield  {journal}
  {\bibinfo  {journal} {Phys. Rev. Lett.}\ }\textbf {\bibinfo {volume} {103}},\
  \bibinfo {pages} {082501} (\bibinfo {year} {2009})},\ \Eprint
  {http://arxiv.org/abs/0905.1873} {arXiv:0905.1873 [nucl-th]} \BibitemShut
  {NoStop}%
\bibitem [{\citenamefont {Jurgenson}\ \emph {et~al.}(2010)\citenamefont
  {Jurgenson}, \citenamefont {Navratil},\ and\ \citenamefont
  {Furnstahl}}]{Jurgenson:2010wy}%
  \BibitemOpen
  \bibfield  {author} {\bibinfo {author} {\bibfnamefont {E.~D.}\ \bibnamefont
  {Jurgenson}}, \bibinfo {author} {\bibfnamefont {P.}~\bibnamefont {Navratil}},
  \ and\ \bibinfo {author} {\bibfnamefont {R.~J.}\ \bibnamefont {Furnstahl}},\
  }\href@noop {} {\  (\bibinfo {year} {2010})},\ \Eprint
  {http://arxiv.org/abs/1011.4085} {arXiv:1011.4085 [nucl-th]} \BibitemShut
  {NoStop}%
\bibitem [{\citenamefont {Hebeler}\ \emph {et~al.}(2010)\citenamefont
  {Hebeler}, \citenamefont {Bogner}, \citenamefont {Furnstahl}, \citenamefont
  {Nogga},\ and\ \citenamefont {Schwenk}}]{Hebeler:2010xb}%
  \BibitemOpen
  \bibfield  {author} {\bibinfo {author} {\bibfnamefont {K.}~\bibnamefont
  {Hebeler}}, \bibinfo {author} {\bibfnamefont {S.~K.}\ \bibnamefont {Bogner}},
  \bibinfo {author} {\bibfnamefont {R.~J.}\ \bibnamefont {Furnstahl}}, \bibinfo
  {author} {\bibfnamefont {A.}~\bibnamefont {Nogga}}, \ and\ \bibinfo {author}
  {\bibfnamefont {A.}~\bibnamefont {Schwenk}},\ }\href@noop {} {\  (\bibinfo
  {year} {2010})},\ \Eprint {http://arxiv.org/abs/1012.3381} {arXiv:1012.3381
  [nucl-th]} \BibitemShut {NoStop}%
\bibitem [{\citenamefont {Navratil}\ \emph
  {et~al.}(2010{\natexlab{a}})\citenamefont {Navratil}, \citenamefont {Roth},\
  and\ \citenamefont {Quaglioni}}]{Navratil:2010jn}%
  \BibitemOpen
  \bibfield  {author} {\bibinfo {author} {\bibfnamefont {P.}~\bibnamefont
  {Navratil}}, \bibinfo {author} {\bibfnamefont {R.}~\bibnamefont {Roth}}, \
  and\ \bibinfo {author} {\bibfnamefont {S.}~\bibnamefont {Quaglioni}},\ }\href
  {\doibase 10.1103/PhysRevC.82.034609} {\bibfield  {journal} {\bibinfo
  {journal} {Phys. Rev.}\ }\textbf {\bibinfo {volume} {C82}},\ \bibinfo {pages}
  {034609} (\bibinfo {year} {2010}{\natexlab{a}})},\ \Eprint
  {http://arxiv.org/abs/1007.0525} {arXiv:1007.0525 [nucl-th]} \BibitemShut
  {NoStop}%
\bibitem [{\citenamefont {Navratil}\ \emph
  {et~al.}(2010{\natexlab{b}})\citenamefont {Navratil}, \citenamefont
  {Quaglioni},\ and\ \citenamefont {Roth}}]{Navratil:2010ey}%
  \BibitemOpen
  \bibfield  {author} {\bibinfo {author} {\bibfnamefont {P.}~\bibnamefont
  {Navratil}}, \bibinfo {author} {\bibfnamefont {S.}~\bibnamefont {Quaglioni}},
  \ and\ \bibinfo {author} {\bibfnamefont {R.}~\bibnamefont {Roth}},\
  }\href@noop {} {\  (\bibinfo {year} {2010}{\natexlab{b}})},\ \Eprint
  {http://arxiv.org/abs/1009.3965} {arXiv:1009.3965 [nucl-th]} \BibitemShut
  {NoStop}%
\bibitem [{\citenamefont {Wiringa}\ \emph {et~al.}(1995)\citenamefont
  {Wiringa}, \citenamefont {Stoks},\ and\ \citenamefont
  {Schiavilla}}]{Wiringa:1994wb}%
  \BibitemOpen
  \bibfield  {author} {\bibinfo {author} {\bibfnamefont {R.~B.}\ \bibnamefont
  {Wiringa}}, \bibinfo {author} {\bibfnamefont {V.~G.~J.}\ \bibnamefont
  {Stoks}}, \ and\ \bibinfo {author} {\bibfnamefont {R.}~\bibnamefont
  {Schiavilla}},\ }\href {\doibase 10.1103/PhysRevC.51.38} {\bibfield
  {journal} {\bibinfo  {journal} {Phys. Rev. C}\ }\textbf {\bibinfo {volume}
  {51}},\ \bibinfo {pages} {38} (\bibinfo {year} {1995})},\ \Eprint
  {http://arxiv.org/abs/nucl-th/9408016} {arXiv:nucl-th/9408016} \BibitemShut
  {NoStop}%
\bibitem [{\citenamefont {Entem}\ and\ \citenamefont
  {Machleidt}(2003)}]{Entem:2003ft}%
  \BibitemOpen
  \bibfield  {author} {\bibinfo {author} {\bibfnamefont {D.~R.}\ \bibnamefont
  {Entem}}\ and\ \bibinfo {author} {\bibfnamefont {R.}~\bibnamefont
  {Machleidt}},\ }\href@noop {} {\bibfield  {journal} {\bibinfo  {journal}
  {Phys. Rev. C}\ }\textbf {\bibinfo {volume} {68}},\ \bibinfo {pages} {041001}
  (\bibinfo {year} {2003})},\ \Eprint {http://arxiv.org/abs/nucl-th/0304018}
  {nucl-th/0304018} \BibitemShut {NoStop}%
\bibitem [{\citenamefont {Epelbaum}\ \emph {et~al.}(2005)\citenamefont
  {Epelbaum}, \citenamefont {Glockle},\ and\ \citenamefont
  {Meissner}}]{Epelbaum:2004fk}%
  \BibitemOpen
  \bibfield  {author} {\bibinfo {author} {\bibfnamefont {E.}~\bibnamefont
  {Epelbaum}}, \bibinfo {author} {\bibfnamefont {W.}~\bibnamefont {Glockle}}, \
  and\ \bibinfo {author} {\bibfnamefont {U.-G.}\ \bibnamefont {Meissner}},\
  }\href@noop {} {\bibfield  {journal} {\bibinfo  {journal} {Nucl. Phys. A}\
  }\textbf {\bibinfo {volume} {747}},\ \bibinfo {pages} {362} (\bibinfo {year}
  {2005})},\ \Eprint {http://arxiv.org/abs/nucl-th/0405048} {nucl-th/0405048}
  \BibitemShut {NoStop}%
\bibitem [{\citenamefont {Nogga}\ \emph {et~al.}(2005)\citenamefont {Nogga},
  \citenamefont {Timmermans},\ and\ \citenamefont {van Kolck}}]{Nogga:2005hy}%
  \BibitemOpen
  \bibfield  {author} {\bibinfo {author} {\bibfnamefont {A.}~\bibnamefont
  {Nogga}}, \bibinfo {author} {\bibfnamefont {R.~G.~E.}\ \bibnamefont
  {Timmermans}}, \ and\ \bibinfo {author} {\bibfnamefont {U.}~\bibnamefont {van
  Kolck}},\ }\href@noop {} {\bibfield  {journal} {\bibinfo  {journal} {Phys.
  Rev. C}\ }\textbf {\bibinfo {volume} {72}},\ \bibinfo {pages} {054006}
  (\bibinfo {year} {2005})},\ \Eprint {http://arxiv.org/abs/nucl-th/0506005}
  {nucl-th/0506005} \BibitemShut {NoStop}%
\bibitem [{\citenamefont {Pavon~Valderrama}\ and\ \citenamefont
  {Arriola}(2006)}]{PavonValderrama:2005wv}%
  \BibitemOpen
  \bibfield  {author} {\bibinfo {author} {\bibfnamefont {M.}~\bibnamefont
  {Pavon~Valderrama}}\ and\ \bibinfo {author} {\bibfnamefont {E.~R.}\
  \bibnamefont {Arriola}},\ }\href {\doibase 10.1103/PhysRevC.74.054001}
  {\bibfield  {journal} {\bibinfo  {journal} {Phys. Rev.}\ }\textbf {\bibinfo
  {volume} {C74}},\ \bibinfo {pages} {054001} (\bibinfo {year} {2006})},\
  \Eprint {http://arxiv.org/abs/nucl-th/0506047} {arXiv:nucl-th/0506047}
  \BibitemShut {NoStop}%
\bibitem [{\citenamefont {Pavon~Valderrama}\ and\ \citenamefont
  {Ruiz~Arriola}(2005)}]{PavonValderrama:2005gu}%
  \BibitemOpen
  \bibfield  {author} {\bibinfo {author} {\bibfnamefont {M.}~\bibnamefont
  {Pavon~Valderrama}}\ and\ \bibinfo {author} {\bibfnamefont {E.}~\bibnamefont
  {Ruiz~Arriola}},\ }\href {\doibase 10.1103/PhysRevC.72.054002} {\bibfield
  {journal} {\bibinfo  {journal} {Phys. Rev.}\ }\textbf {\bibinfo {volume}
  {C72}},\ \bibinfo {pages} {054002} (\bibinfo {year} {2005})},\ \Eprint
  {http://arxiv.org/abs/nucl-th/0504067} {arXiv:nucl-th/0504067} \BibitemShut
  {NoStop}%
\bibitem [{\citenamefont {Pavon~Valderrama}\ and\ \citenamefont
  {Ruiz~Arriola}(2006)}]{PavonValderrama:2005uj}%
  \BibitemOpen
  \bibfield  {author} {\bibinfo {author} {\bibfnamefont {M.}~\bibnamefont
  {Pavon~Valderrama}}\ and\ \bibinfo {author} {\bibfnamefont {E.}~\bibnamefont
  {Ruiz~Arriola}},\ }\href {\doibase 10.1103/PhysRevC.74.064004} {\bibfield
  {journal} {\bibinfo  {journal} {Phys. Rev.}\ }\textbf {\bibinfo {volume}
  {C74}},\ \bibinfo {pages} {064004} (\bibinfo {year} {2006})},\ \Eprint
  {http://arxiv.org/abs/nucl-th/0507075} {arXiv:nucl-th/0507075} \BibitemShut
  {NoStop}%
\bibitem [{\citenamefont {Pavon~Valderrama}()}]{Valderrama:2009ei}%
  \BibitemOpen
  \bibfield  {author} {\bibinfo {author} {\bibfnamefont {M.}~\bibnamefont
  {Pavon~Valderrama}},\ }\href@noop {} {\ }\Eprint
  {http://arxiv.org/abs/0912.0699} {arXiv:0912.0699 [nucl-th]} \BibitemShut
  {NoStop}%
\bibitem [{\citenamefont {Yang}\ \emph {et~al.}(2009)\citenamefont {Yang},
  \citenamefont {Elster},\ and\ \citenamefont {Phillips}}]{Yang:2009fm}%
  \BibitemOpen
  \bibfield  {author} {\bibinfo {author} {\bibfnamefont {C.~J.}\ \bibnamefont
  {Yang}}, \bibinfo {author} {\bibfnamefont {C.}~\bibnamefont {Elster}}, \ and\
  \bibinfo {author} {\bibfnamefont {D.~R.}\ \bibnamefont {Phillips}},\
  }\href@noop {} {\bibfield  {journal} {\bibinfo  {journal} {PoS}\ }\textbf
  {\bibinfo {volume} {CD09}},\ \bibinfo {pages} {064} (\bibinfo {year}
  {2009})},\ \Eprint {http://arxiv.org/abs/0909.5414} {arXiv:0909.5414
  [nucl-th]} \BibitemShut {NoStop}%
\bibitem [{\citenamefont {Machleidt}\ \emph {et~al.}(2010)\citenamefont
  {Machleidt}, \citenamefont {Liu}, \citenamefont {Entem},\ and\ \citenamefont
  {Arriola}}]{Machleidt:2009bh}%
  \BibitemOpen
  \bibfield  {author} {\bibinfo {author} {\bibfnamefont {R.}~\bibnamefont
  {Machleidt}}, \bibinfo {author} {\bibfnamefont {P.}~\bibnamefont {Liu}},
  \bibinfo {author} {\bibfnamefont {D.~R.}\ \bibnamefont {Entem}}, \ and\
  \bibinfo {author} {\bibfnamefont {E.~R.}\ \bibnamefont {Arriola}},\ }\href
  {\doibase 10.1103/PhysRevC.81.024001} {\bibfield  {journal} {\bibinfo
  {journal} {Phys. Rev.}\ }\textbf {\bibinfo {volume} {C81}},\ \bibinfo {pages}
  {024001} (\bibinfo {year} {2010})},\ \Eprint {http://arxiv.org/abs/0910.3942}
  {arXiv:0910.3942 [nucl-th]} \BibitemShut {NoStop}%
\bibitem [{\citenamefont {Phillips}(2010)}]{Phillips:2010rv}%
  \BibitemOpen
  \bibfield  {author} {\bibinfo {author} {\bibfnamefont {D.~R.}\ \bibnamefont
  {Phillips}},\ }\href@noop {} {\  (\bibinfo {year} {2010})},\ \Eprint
  {http://arxiv.org/abs/1010.0622} {arXiv:1010.0622 [nucl-th]} \BibitemShut
  {NoStop}%
\bibitem [{\citenamefont {Glazek}\ and\ \citenamefont
  {Perry}(2008)}]{Glazek:2008pg}%
  \BibitemOpen
  \bibfield  {author} {\bibinfo {author} {\bibfnamefont {S.~D.}\ \bibnamefont
  {Glazek}}\ and\ \bibinfo {author} {\bibfnamefont {R.~J.}\ \bibnamefont
  {Perry}},\ }\href {\doibase 10.1103/PhysRevD.78.045011} {\bibfield  {journal}
  {\bibinfo  {journal} {Phys. Rev. D}\ }\textbf {\bibinfo {volume} {78}},\
  \bibinfo {pages} {045011} (\bibinfo {year} {2008})},\ \Eprint
  {http://arxiv.org/abs/0803.2911} {arXiv:0803.2911 [nucl-th]} \BibitemShut
  {NoStop}%
\bibitem [{\citenamefont {Wegner}(1994)}]{Wegner:1994}%
  \BibitemOpen
  \bibfield  {author} {\bibinfo {author} {\bibfnamefont {F.}~\bibnamefont
  {Wegner}},\ }\href@noop {} {\bibfield  {journal} {\bibinfo  {journal} {Ann.\
  Phys.\ (Leipzig)}\ }\textbf {\bibinfo {volume} {3}},\ \bibinfo {pages} {77}
  (\bibinfo {year} {1994})}\BibitemShut {NoStop}%
\bibitem [{\citenamefont {Kehrein}(2006)}]{Kehrein:2006}%
  \BibitemOpen
  \bibfield  {author} {\bibinfo {author} {\bibfnamefont {S.}~\bibnamefont
  {Kehrein}},\ }\href@noop {} {\emph {\bibinfo {title} {The Flow Equation
  Approach to Many-Particle Systems}}}\ (\bibinfo  {publisher} {Springer},\
  \bibinfo {address} {Berlin},\ \bibinfo {year} {2006})\BibitemShut {NoStop}%
\end{thebibliography}%
\end{document}